%%%
%%%  Feb. 9, 2005
%%%
%%% Feb. 11, 2005
%%% Last edits by gm, Feb. 12, 2005
%%%
%%% Version 2 by gm, Feb. 13, 2005
%%%
%%% Version 3 by ad, Feb. 14, 2005
%%%
%%% Version 4 by gm, Feb. 14, 20:00 GMT.
%%%
%%% Version 5 by ad, Feb 15, 1800 GMT
%%%
%%% Version 5.1 by ad, Feb 16, 1600 GMT
%%%
%%% Version 6 by gm, Feb. 16, 14:30 GMT
%%%
%%% V8 by boris
%%%
%%% V9 by gm, 19:40 GMT
%%%
%%% V11 by ad Feb 17 14:30GMT
%%%
%%% V12 by gm, 16:20 GMT, combining versions of ad and bp.
%%%
%%%

%%% Revisions by gm, March 9, 2005,   revisions March 14, 2005
%%% Revisions March 20, 2005
%%%
%%% Substantial revisions by gm, March 22, 2005; April 4, 2005
%%% Atish's conclusions April 10
%%% More revisions by gm, April 18, April 19
%%%
%%%
%%%  More revisions, April 20,....
%%%
%%% Final revisions, April 21
%%%
%%% Further revisions to fix mistake, April 29.
%%%
%%% V3.2 by fd, May 1
%%% Revised by gm, May 4
%%%
%%% Revised by gm, May 12
%%%
%%%

\input harvmac.tex

\def\vt#1#2#3{ {\vartheta[{#1 \atop  #2}](#3\vert \tau)} }

\def\makeblankbox#1#2{\hbox{\lower\dp0\vbox{\hidehrule{#1}{#2}%
   \kern -#1% overlap rules
   \hbox to \wd0{\hidevrule{#1}{#2}%
      \raise\ht0\vbox to #1{}% vrule height
      \lower\dp0\vtop to #1{}% vrule depth
      \hfil\hidevrule{#2}{#1}}%
   \kern-#1\hidehrule{#2}{#1}}}%
}%
\def\hidehrule#1#2{\kern-#1\hrule height#1 depth#2 \kern-#2}%
\def\hidevrule#1#2{\kern-#1{\dimen0=#1\advance\dimen0 by #2\vrule
    width\dimen0}\kern-#2}%
\def\openbox{\ht0=1.2mm \dp0=1.2mm \wd0=2.4mm  \raise 2.75pt
\makeblankbox {.25pt} {.25pt}  }

\def\bun#1/#2{\leavevmode
   \kern.1em \raise .5ex \hbox{\the\scriptfont0 #1}%
   \kern-.1em $/$%
   \kern-.15em \lower .25ex \hbox{\the\scriptfont0 #2}%
}

\def\opensquare{\ht0=3.4mm \dp0=3.4mm \wd0=6.8mm  \raise 2.7pt
\makeblankbox {.25pt} {.25pt}  }

%%%%%%%%%%%%%%%%%%%%%%%

\def\sector#1#2{\ {\scriptstyle #1}\hskip 1mm
\mathop{\opensquare}\limits_{\lower
1mm\hbox{$\scriptstyle#2$}}\hskip 1mm}

\def\tsector#1#2{\ {\scriptstyle #1}\hskip 1mm
\mathop{\opensquare}\limits_{\lower
1mm\hbox{$\scriptstyle#2$}}^\sim\hskip 1mm}
%%%
%%%

\font\ticp=cmcsc10

\def\la{{\lambda}}
\def\IZ{{\bf Z}}
\def\IC{{\bf C}}
\def\IP{{\bf P}}
\def\IQ{{\bf Q}}
\def\IR{{\bf R}}
\def\CM{{\cal M}}
\def\CN{{\cal N}}
\def\CQ{{\cal Q}}
\def\CX{{\cal X}}
\def\CZ{{\cal Z}}
\def\mod{{\rm mod}}
\def\det{{\rm det}}
\def\p{\partial}

\def\vt#1#2#3{ {\vartheta[{#1 \atop  #2}](#3\vert \tau)} }

\def\im{{\rm Im}}

\def\frac#1#2{{#1\over#2}}
\def\ZZ{\IZ}

%% Something to deal with sub-sub-sections

\def\unlockat{\catcode`\@=11}
\def\lockat{\catcode`\@=12}

\unlockat
%%% Something to deal with sub-sub-sections

\def\newsec#1{\global\advance\secno by1\message{(\the\secno. #1)}
\global\subsecno=0\global\subsubsecno=0\eqnres@t\noindent
{\bf\the\secno. #1}
\writetoca{{\secsym} {#1}}\par\nobreak\medskip\nobreak}
\global\newcount\subsecno \global\subsecno=0
\def\subsec#1{\global\advance\subsecno
by1\message{(\secsym\the\subsecno. #1)}
\ifnum\lastpenalty>9000\else\bigbreak\fi\global\subsubsecno=0
\noindent{\it\secsym\the\subsecno. #1}
\writetoca{\string\quad {\secsym\the\subsecno.} {#1}}
\par\nobreak\medskip\nobreak}
\global\newcount\subsubsecno \global\subsubsecno=0
\def\subsubsec#1{\global\advance\subsubsecno by1
\message{(\secsym\the\subsecno.\the\subsubsecno. #1)}
\ifnum\lastpenalty>9000\else\bigbreak\fi
\noindent\quad{\secsym\the\subsecno.\the\subsubsecno.}{#1}
\writetoca{\string\qquad{\secsym\the\subsecno.\the\subsubsecno.}{#1}}
\par\nobreak\medskip\nobreak}

\def\subsubseclab#1{\DefWarn#1\xdef
#1{\noexpand\hyperref{}{subsubsection}%
{\secsym\the\subsecno.\the\subsubsecno}%
{\secsym\the\subsecno.\the\subsubsecno}}%
\writedef{#1\leftbracket#1}\wrlabeL{#1=#1}}% Macros for boxes
\lockat

%%%%%%%%%%%%%%%%%%%%%%%%%%%%%%%%%%%%%%%%%%%%%%%%
%%%%%%%%%%%%%%%%%%%%%%%%%%%%%%%%%%%%%%%%%%%%%%
%%%%%%%%%%%%%%%%%%%%%%%%%% REFERENCES
%%%%%%%%%%%%%%%%%%%%%%%%%%%%%%%%%%%%%%%%%%%
%%%%%%%%%%%%%%%%%%%%%%%%%%%%%%%%%%%%%%%%%%%%%%%

%\BekensteinUR
\lref\BekensteinUR{ J.~D.~Bekenstein,
``Black Holes And Entropy,''
Phys.\ Rev.\ D {\bf 7}, 2333 (1973).
%%CITATION = PHRVA,D7,2333;%%
}

%\BekensteinAX
\lref\BekensteinAX{ J.~D.~Bekenstein,
``Generalized Second Law Of
Thermodynamics In Black Hole Physics,'' Phys.\ Rev.\ D {\bf 9},
3292 (1974).
%%CITATION = PHRVA,D9,3292;%%
}

%\HawkingSW
\lref\HawkingSW{ S.~W.~Hawking,
``Particle Creation By Black
Holes,'' Commun.\ Math.\ Phys.\  {\bf 43}, 199 (1975).
%%CITATION = CMPHA,43,199;%%
}

%\FerraraIH
\lref\FerraraIH{ S.~Ferrara, R.~Kallosh and A.~Strominger,
``N=2 extremal black holes,''
Phys.\ Rev.\ D {\bf 52}, 5412 (1995) [arXiv:hep-th/9508072].
%%CITATION = HEP-TH 9508072;%%
}

%\StromingerKF
\lref\StromingerKF{ A.~Strominger,
``Macroscopic Entropy of $N=2$ Extremal Black Holes,''
Phys.\ Lett.\ B {\bf 383}, 39 (1996) [arXiv:hep-th/9602111].
%%CITATION = HEP-TH 9602111;%%
}
%\MohauptGC
\lref\MohauptGC{ T.~Mohaupt,
``Black holes in supergravity and string theory,''
Class.\ Quant.\ Grav.\  {\bf 17}, 3429 (2000)
[arXiv:hep-th/0004098].
%%CITATION = HEP-TH 0004098;%%
}

%\CardosoXF
\lref\CardosoXF{ G.~L.~Cardoso, B.~de Wit, J.~Kappeli and
T.~Mohaupt,
%``Asymptotic degeneracy of dyonic N = 4 string states and black hole
%entropy,''
arXiv:hep-th/0412287.
%%CITATION = HEP-TH 0412287;%%
}

%\WaldNT
\lref\WaldNT{ R.~M.~Wald,
``Black hole entropy in the Noether
charge,'' Phys.\ Rev.\ D {\bf 48}, 3427 (1993)
[arXiv:gr-qc/9307038].
%%CITATION = GR-QC 9307038;%%
}

%\IyerYS
\lref\IyerYS{ V.~Iyer and R.~M.~Wald,
``Some properties of Noether
charge and a proposal for dynamical black hole entropy,'' Phys.\
Rev.\ D {\bf 50}, 846 (1994) [arXiv:gr-qc/9403028].
%%CITATION = GR-QC 9403028;%%
}

%\LopesCardosoWT
\lref\LopesCardosoWT{ G.~Lopes Cardoso, B.~de Wit and T.~Mohaupt,
``Corrections to macroscopic supersymmetric black-hole entropy,''
Phys.\ Lett.\ B {\bf 451}, 309 (1999) [arXiv:hep-th/9812082].
%%CITATION = HEP-TH 9812082;%%
}

%\LopesCardosoCV
\lref\LopesCardosoCV{ G.~Lopes Cardoso, B.~de Wit and T.~Mohaupt,
``Deviations from the area law for supersymmetric black holes,''
Fortsch.\ Phys.\  {\bf 48}, 49 (2000) [arXiv:hep-th/9904005].
%%CITATION = HEP-TH 9904005;%%
}

%\LopesCardosoXN
\lref\LopesCardosoXN{ G.~Lopes Cardoso, B.~de Wit and T.~Mohaupt,
``Area law corrections from state counting and supergravity,''
Class.\ Quant.\ Grav.\  {\bf 17}, 1007 (2000)
[arXiv:hep-th/9910179].
%%CITATION = HEP-TH 9910179;%%
}

%\LopesCardosoQM
\lref\LopesCardosoQM{ G.~Lopes Cardoso, B.~de Wit, J.~Kappeli and
T.~Mohaupt, ``Stationary BPS solutions in N = 2 supergravity with
$R^2$ interactions,'' JHEP {\bf 0012}, 019 (2000)
[arXiv:hep-th/0009234].
%%CITATION = HEP-TH 0009234;%%
}

%\CardosoXF
\lref\CardosoXF{ G.~L.~Cardoso, B.~de Wit, J.~Kappeli and
T.~Mohaupt, ``Asymptotic degeneracy of dyonic N = 4 string states
and black hole entropy,'' arXiv:hep-th/0412287.
%%CITATION = HEP-TH 0412287;%%
}

%\LopesCardosoUR
\lref\LopesCardosoUR{ G.~Lopes Cardoso, B.~de Wit and T.~Mohaupt,
``Macroscopic entropy formulae and non-holomorphic corrections for
supersymmetric black holes,'' Nucl.\ Phys.\ B {\bf 567}, 87 (2000)
[arXiv:hep-th/9906094].
%%CITATION = HEP-TH 9906094;%%
}

%\SenIN
\lref\SenIN{ A.~Sen,
``Extremal black holes and elementary string
states,'' Mod.\ Phys.\ Lett.\ A {\bf 10}, 2081 (1995)
[arXiv:hep-th/9504147].
%%CITATION = HEP-TH 9504147;%%
}

%\SenIS
\lref\SenIS{ A.~Sen,
``Black holes and elementary string states in
N = 2 supersymmetric string theories,''
JHEP {\bf 9802}, 011
(1998) [arXiv:hep-th/9712150].
%%CITATION = HEP-TH 9712150;%%
}

%\DabholkarJT
\lref\DabholkarJT{ A.~Dabholkar and J.~A.~Harvey,
``Nonrenormalization Of The Superstring Tension,'' Phys.\ Rev.\
Lett.\  {\bf 63}, 478 (1989).
%%CITATION = PRLTA,63,478;%%
}

%\DabholkarYF
\lref\DabholkarYF{ A.~Dabholkar, G.~W.~Gibbons, J.~A.~Harvey and
F.~Ruiz Ruiz,
``Superstrings And Solitons,'' Nucl.\ Phys.\ B {\bf
340}, 33 (1990).
%%CITATION = NUPHA,B340,33;%%
}

%\SenEB
\lref\SenEB{ A.~Sen,
``Black hole solutions in heterotic string
theory on a torus,'' Nucl.\ Phys.\ B {\bf 440}, 421 (1995)
[arXiv:hep-th/9411187].
%%CITATION = HEP-TH 9411187;%%
}

%\DabholkarNC
\lref\DabholkarNC{ A.~Dabholkar, J.~P.~Gauntlett, J.~A.~Harvey and
D.~Waldram,
``Strings as Solitons \& Black Holes as Strings,''
Nucl.\ Phys.\ B {\bf 474}, 85 (1996) [arXiv:hep-th/9511053].
%%CITATION = HEP-TH 9511053;%%
}

%\LopesCardosoUR
\lref\LopesCardosoUR{ G.~Lopes Cardoso, B.~de Wit and T.~Mohaupt,
``Macroscopic entropy formulae and non-holomorphic corrections for
supersymmetric black holes,'' Nucl.\ Phys.\ B {\bf 567}, 87 (2000)
[arXiv:hep-th/9906094].
%%CITATION = HEP-TH 9906094;%%
}

%\BershadskyCX
\lref\BershadskyCX{ M.~Bershadsky, S.~Cecotti, H.~Ooguri and
C.~Vafa, ``Kodaira-Spencer theory of gravity and exact results for
quantum string  amplitudes,'' Commun.\ Math.\ Phys.\  {\bf 165},
311 (1994) [arXiv:hep-th/9309140].
%%CITATION = HEP-TH 9309140;%%
}

%\BershadskyTA
\lref\BershadskyTA{ M.~Bershadsky, S.~Cecotti, H.~Ooguri and
C.~Vafa,
``Holomorphic anomalies in topological field theories,''
Nucl.\ Phys.\ B {\bf 405}, 279 (1993) [arXiv:hep-th/9302103].
%%CITATION = HEP-TH 9302103;%%
}

%\AntoniadisZE
\lref\AntoniadisZE{ I.~Antoniadis, E.~Gava, K.~S.~Narain and
T.~R.~Taylor,
``Topological amplitudes in string theory,'' Nucl.\
Phys.\ B {\bf 413}, 162 (1994) [arXiv:hep-th/9307158].
%%CITATION = HEP-TH 9307158;%%
}

%\ChaudhuriFK
\lref\ChaudhuriFK{ S.~Chaudhuri, G.~Hockney and J.~D.~Lykken,
``Maximally supersymmetric string theories in $D < 10$,'' Phys.\
Rev.\ Lett.\  {\bf 75}, 2264 (1995) [arXiv:hep-th/9505054].
%%CITATION = HEP-TH 9505054;%%
}

%\ChaudhuriBF
\lref\ChaudhuriBF{ S.~Chaudhuri and J.~Polchinski, ``Moduli space
of CHL strings,'' Phys.\ Rev.\ D {\bf 52}, 7168 (1995)
[arXiv:hep-th/9506048].
%%CITATION = HEP-TH 9506048;%%
}

%\HarveyTS
\lref\HarveyTS{ J.~A.~Harvey and G.~W.~Moore, ``Exact
gravitational threshold correction in the FHSV model,'' Phys.\
Rev.\ D {\bf 57}, 2329 (1998) [arXiv:hep-th/9611176].
%%CITATION = HEP-TH 9611176;%%
}

%\AspinwallVK
\lref\AspinwallVK{ P.~S.~Aspinwall and J.~Louis, ``On the Ubiquity
of K3 Fibrations in String Duality,'' Phys.\ Lett.\ B {\bf 369},
233 (1996) [arXiv:hep-th/9510234].
%%CITATION = HEP-TH 9510234;%%
}

%\HarveyFQ
\lref\HarveyFQ{ J.~A.~Harvey and G.~W.~Moore, ``Algebras, BPS
States, and Strings,'' Nucl.\ Phys.\ B {\bf 463}, 315 (1996)
[arXiv:hep-th/9510182].
%%CITATION = HEP-TH 9510182;%%
}

%\DijkgraafFQ
\lref\DijkgraafFQ{ R.~Dijkgraaf, J.~M.~Maldacena, G.~W.~Moore and
E.~Verlinde, ``A black hole Farey tail,'' arXiv:hep-th/0005003.
%%CITATION = HEP-TH 0005003;%%
}

%\DabholkarYR
\lref\DabholkarYR{ A.~Dabholkar, ``Exact counting of black hole
microstates,'' arXiv:hep-th/0409148.
%%CITATION = HEP-TH 0409148;%%
}

%\DabholkarDQ
\lref\DabholkarDQ{ A.~Dabholkar, R.~Kallosh and A.~Maloney, ``A
stringy cloak for a classical singularity,'' JHEP {\bf 0412}, 059
(2004) [arXiv:hep-th/0410076].
%%CITATION = HEP-TH 0410076;%%
}

%\OoguriZV
\lref\OoguriZV{ H.~Ooguri, A.~Strominger and C.~Vafa, ``Black hole
attractors and the topological string,'' Phys.\ Rev.\ D {\bf 70},
106007 (2004) [arXiv:hep-th/0405146].
%%CITATION = HEP-TH 0405146;%%
}

%\MaldacenaDE
\lref\MaldacenaDE{ J.~M.~Maldacena, A.~Strominger and E.~Witten,
``Black hole entropy in M-theory,'' JHEP {\bf 9712}, 002 (1997)
[arXiv:hep-th/9711053].
%%CITATION = HEP-TH 9711053;%%
}

%\HarveyTS
\lref\HarveyTS{ J.~A.~Harvey and G.~W.~Moore, ``Exact
gravitational threshold correction in the FHSV model,'' Phys.\
Rev.\ D {\bf 57}, 2329 (1998) [arXiv:hep-th/9611176].
%%CITATION = HEP-TH 9611176;%%
}

%\AspinwallVK
\lref\AspinwallVK{ P.~S.~Aspinwall and J.~Louis, ``On the Ubiquity
of K3 Fibrations in String Duality,'' Phys.\ Lett.\ B {\bf 369},
233 (1996) [arXiv:hep-th/9510234].
%%CITATION = HEP-TH 9510234;%%
}

%\HarveyFQ
\lref\HarveyFQ{ J.~A.~Harvey and G.~W.~Moore, ``Algebras, BPS
States, and Strings,'' Nucl.\ Phys.\ B {\bf 463}, 315 (1996)
[arXiv:hep-th/9510182].
%%CITATION = HEP-TH 9510182;%%
}

%\BrunnerSK
\lref\BrunnerSK{ I.~Brunner, J.~Distler and R.~Mahajan, ``Return
of the torsion D-branes,'' Adv.\ Theor.\ Math.\ Phys.\  {\bf 5},
311 (2002) [arXiv:hep-th/0106262].
%%CITATION = HEP-TH 0106262;%%
}

%\FerraraYX
\lref\FerraraYX{ S.~Ferrara, J.~A.~Harvey, A.~Strominger and
C.~Vafa, ``Second quantized mirror symmetry,'' Phys.\ Lett.\ B
{\bf 361}, 59 (1995) [arXiv:hep-th/9505162].
%%CITATION = HEP-TH 9505162;%%
}

%\MillerAG
\lref\MillerAG{ S.~D.~Miller and G.~W.~Moore, ``Landau-Siegel
zeroes and black hole entropy,'' arXiv:hep-th/9903267.
%%CITATION = HEP-TH 9903267;%%
}

%\KiritsisHJ
\lref\KiritsisHJ{ E.~Kiritsis, ``Introduction to superstring
theory,'' arXiv:hep-th/9709062.
%%CITATION = HEP-TH 9709062;%%
}

%\VafaQA
\lref\VafaQA{ C.~Vafa, ``Two dimensional Yang-Mills, black holes
and topological strings,'' arXiv:hep-th/0406058.
%%CITATION = HEP-TH 0406058;%%
}
%\AganagicJS
\lref\AganagicJS{ M.~Aganagic, H.~Ooguri, N.~Saulina and C.~Vafa,
``Black holes, q-deformed 2d Yang-Mills, and non-perturbative
topological strings,'' arXiv:hep-th/0411280.
%%CITATION = HEP-TH 0411280;%%
}

%\SenDP
\lref\SenDP{ A.~Sen, ``How does a fundamental string stretch its
horizon?,'' arXiv:hep-th/0411255.
%%CITATION = HEP-TH 0411255;%%
}

%\HubenyJI
\lref\HubenyJI{ V.~Hubeny, A.~Maloney and M.~Rangamani,
``String-corrected black holes,'' arXiv:hep-th/0411272.
%%CITATION = HEP-TH 0411272;%%
}

%\StromingerSH
\lref\StromingerSH{ A.~Strominger and C.~Vafa, ``Microscopic
Origin of the Bekenstein-Hawking Entropy,'' Phys.\ Lett.\ B {\bf
379}, 99 (1996) [arXiv:hep-th/9601029].
%%CITATION = HEP-TH 9601029;%%
}

%\DamourKW
\lref\DamourKW{ T.~Damour, ``The entropy of black holes: A
primer,'' arXiv:hep-th/0401160.
%%CITATION = HEP-TH 0401160;%%
}

%\MarinoPG
\lref\MarinoPG{ M.~Marino and G.~W.~Moore, ``Counting higher genus
curves in a Calabi-Yau manifold,'' Nucl.\ Phys.\ B {\bf 543}, 592
(1999) [arXiv:hep-th/9808131].
%%CITATION = HEP-TH 9808131;%%
}

%\KlemmKM
\lref\KlemmKM{ A.~Klemm, M.~Kreuzer, E.~Riegler and
E.~Scheidegger, ``Topological string amplitudes, complete
intersection Calabi-Yau spaces and threshold corrections,''
arXiv:hep-th/0410018.
%%CITATION = HEP-TH 0410018;%%
}

%\VerlindeCK
\lref\VerlindeCK{ E.~Verlinde, ``Attractors and the holomorphic
anomaly,'' arXiv:hep-th/0412139.
%%CITATION = HEP-TH 0412139;%%
}

%\MooreAR
\lref\MooreAR{ G.~W.~Moore, ``String duality, automorphic forms,
and generalized Kac-Moody algebras,'' Nucl.\ Phys.\ Proc.\ Suppl.\
{\bf 67}, 56 (1998) [arXiv:hep-th/9710198].
%%CITATION = HEP-TH 9710198;%%
}

%\HosonoAV
\lref\HosonoAV{ S.~Hosono, A.~Klemm and S.~Theisen, ``Lectures on
mirror symmetry,'' arXiv:hep-th/9403096.
%%CITATION = HEP-TH 9403096;%%
}

%\HosonoAX
\lref\HosonoAX{ S.~Hosono, A.~Klemm, S.~Theisen and S.~T.~Yau,
``Mirror symmetry, mirror map and applications to complete
intersection Calabi-Yau spaces,'' Nucl.\ Phys.\ B {\bf 433}, 501
(1995) [arXiv:hep-th/9406055].
%%CITATION = HEP-TH 9406055;%%
}

%\SenPU
\lref\SenPU{ A.~Sen, ``Black Holes, Elementary Strings and
Holomorphic Anomaly,'' arXiv:hep-th/0502126.
%%CITATION = HEP-TH 0502126;%%
}

\lref\DDMP{A.~Dabholkar, F.~Denef, G.~W.~Moore and B.~Pioline,
``Precision counting of small black holes,''
arXiv:hep-th/0507014.
%%CITATION = HEP-TH 0507014;%%
}

%\PiolineVI
\lref\PiolineVI{
B.~Pioline,
``BPS black hole degeneracies and minimal automorphic representations,''
arXiv:hep-th/0506228.
%%CITATION = HEP-TH 0506228;%%
}

%\WittenXJ
\lref\WittenXJ{
E.~Witten,
``Topological Sigma Models,''
Commun.\ Math.\ Phys.\  {\bf 118}, 411 (1988).
%%CITATION = CMPHA,118,411;%%
}

%\WittenZZ
\lref\WittenZZ{
E.~Witten,
``Mirror manifolds and topological field theory,''
arXiv:hep-th/9112056.
%%CITATION = HEP-TH 9112056;%%
}

%\FerraraQD
\lref\FerraraQD{
S.~Ferrara, C.~A.~Savoy and L.~Girardello,
``Spin Sum Rules In Extended Supersymmetry,''
Phys.\ Lett.\ B {\bf 105}, 363 (1981).
%%CITATION = PHLTA,B105,363;%%
}

%\CecottiQH
\lref\CecottiQH{ S.~Cecotti, P.~Fendley, K.~A.~Intriligator and
C.~Vafa,
``A New supersymmetric index,''
Nucl.\ Phys.\ B {\bf 386}, 405 (1992) [arXiv:hep-th/9204102].
%%CITATION = HEP-TH 9204102;%%
}

%\GregoriHI
\lref\GregoriHI{
A.~Gregori, E.~Kiritsis, C.~Kounnas, N.~A.~Obers, P.~M.~Petropoulos and
B.~Pioline,
``$R^2$ corrections and non-perturbative dualities of N = 4 string ground
states,''
Nucl.\ Phys.\ B {\bf 510}, 423 (1998)
[arXiv:hep-th/9708062].
%%CITATION = HEP-TH 9708062;%%
}

%\SchwarzBJ
\lref\SchwarzBJ{
J.~H.~Schwarz and A.~Sen,
``Type IIA dual of the six-dimensional CHL compactification,''
Phys.\ Lett.\ B {\bf 357}, 323 (1995)
[arXiv:hep-th/9507027].
%%CITATION = HEP-TH 9507027;%%
}

%\FerraraIH
\lref\FerraraIH{
S.~Ferrara, R.~Kallosh and A.~Strominger,
``N=2 extremal black holes,''
Phys.\ Rev.\ D {\bf 52}, 5412 (1995)
[arXiv:hep-th/9508072].
%%CITATION = HEP-TH 9508072;%%
}

%\LercheZY
\lref\LercheZY{
W.~Lerche,
``Elliptic Index And Superstring Effective Actions,''
Nucl.\ Phys.\ B {\bf 308}, 102 (1988).
%%CITATION = NUPHA,B308,102;%%
}

%\LercheZZ
\lref\LercheZZ{
W.~Lerche, A.~N.~Schellekens and N.~P.~Warner,
``Ghost Triality And Superstring Partition Functions,''
Phys.\ Lett.\ B {\bf 214}, 41 (1988).
%%CITATION = PHLTA,B214,41;%%
}

%\GopakumarJQ
\lref\GopakumarJQ{
  R.~Gopakumar and C.~Vafa,
  ``M-theory and topological strings. II,''
  arXiv:hep-th/9812127.
  %%CITATION = HEP-TH 9812127;%%
}

%\BakMT
\lref\BakMT{
  D.~Bak, S.~Kim and S.~J.~Rey,
  ``Exactly soluble BPS black holes in higher curvature N = 2 supergravity,''
  arXiv:hep-th/0501014.
  %%CITATION = HEP-TH 0501014;%%
}

%\MohauptMJ
\lref\MohauptMJ{
  T.~Mohaupt,
  ``Black hole entropy, special geometry and strings,''
  Fortsch.\ Phys.\  {\bf 49}, 3 (2001)
  [arXiv:hep-th/0007195].
  %%CITATION = HEP-TH 0007195;%%
}

%\deWitFZ
\lref\deWitFZ{
  B.~de Wit,
  ``Introduction to black hole entropy and supersymmetry,''
  arXiv:hep-th/0503211.
  %%CITATION = HEP-TH 0503211;%%
}

%\DixonPC
\lref\DixonPC{
  L.~J.~Dixon, V.~Kaplunovsky and J.~Louis,
  ``Moduli Dependence Of String Loop Corrections To Gauge Coupling Constants,''
  Nucl.\ Phys.\ B {\bf 355}, 649 (1991).
  %%CITATION = NUPHA,B355,649;%%
}

%\deWitZG
\lref\deWitZG{
  B.~de Wit, V.~Kaplunovsky, J.~Louis and D.~Lust,
  ``Perturbative couplings of vector multiplets in N=2 heterotic string
  vacua,''
  Nucl.\ Phys.\ B {\bf 451}, 53 (1995)
  [arXiv:hep-th/9504006].
  %%CITATION = HEP-TH 9504006;%%
}

%\AntoniadisCT
\lref\AntoniadisCT{
  I.~Antoniadis, S.~Ferrara, E.~Gava, K.~S.~Narain and T.~R.~Taylor,
  ``Perturbative prepotential and monodromies in N=2 heterotic superstring,''
  Nucl.\ Phys.\ B {\bf 447}, 35 (1995)
  [arXiv:hep-th/9504034].
  %%CITATION = HEP-TH 9504034;%%
}

\lref\millergelbart{S.S. Gelbart and S.D. Miller, ``Riemann's zeta function
and beyond,''  Bull. Amer. Math. Soc. {\bf 41} 59}

%\DijkgraafBP
\lref\DijkgraafBP{
  R.~Dijkgraaf, R.~Gopakumar, H.~Ooguri and C.~Vafa,
  ``Baby Universes in String Theory,''
  arXiv:hep-th/0504221.
  %%CITATION = HEP-TH 0504221;%%
}

%\VafaQA
\lref\VafaQA{
  C.~Vafa,
  ``Two dimensional Yang-Mills, black holes and topological strings,''
  arXiv:hep-th/0406058.
  %%CITATION = HEP-TH 0406058;%%
}

%\AganagicJS
\lref\AganagicJS{
  M.~Aganagic, H.~Ooguri, N.~Saulina and C.~Vafa,
  ``Black holes, q-deformed 2d Yang-Mills, and non-perturbative topological
  strings,''
  arXiv:hep-th/0411280.
  %%CITATION = HEP-TH 0411280;%%
}

%\MinasianQN
\lref\MinasianQN{
  R.~Minasian, G.~W.~Moore and D.~Tsimpis,
  ``Calabi-Yau black holes and (0,4) sigma models,''
  Commun.\ Math.\ Phys.\  {\bf 209}, 325 (2000)
  [arXiv:hep-th/9904217].
  %%CITATION = HEP-TH 9904217;%%
}

%\HenningsonJZ
\lref\HenningsonJZ{
  M.~Henningson and G.~W.~Moore,
  ``Threshold corrections in K(3) x T(2) heterotic string  compactifications,''
  Nucl.\ Phys.\ B {\bf 482}, 187 (1996)
  [arXiv:hep-th/9608145].
  %%CITATION = HEP-TH 9608145;%%
}

%\deWitFZ
\lref\deWitFZ{
  B.~de Wit,
  ``Introduction to black hole entropy and supersymmetry,''
  arXiv:hep-th/0503211.
  %%CITATION = HEP-TH 0503211;%%
}

%\SeibergXZ
\lref\SeibergXZ{
  N.~Seiberg and E.~Witten,
  ``The D1/D5 system and singular CFT,''
  JHEP {\bf 9904}, 017 (1999)
  [arXiv:hep-th/9903224].
  %%CITATION = HEP-TH 9903224;%%
}

%%%%%%%%%%%%%%%%%%%%%%%%%%%%%%%%%%%%%%%%%%%%%%%
%%%%%%%%%%%%%%%%%%%%%%%%%%%%%%%%%%%%%%%%%%%%%%%%

%-------------------
% title page
%-------------------
%
\Title{\vbox{\baselineskip12pt \hbox{hep-th/0502157}
\hbox{LPTHE-05-04}\hbox{LPTENS-05-09} \hbox{TIFR-TH-05-07}} }
{\vbox{\centerline{Exact and Asymptotic Degeneracies of }
\bigskip
\centerline{Small Black Holes  }}} \centerline{\ticp Atish
Dabholkar$^{1}$, Frederik Denef$^{2}$, Gregory W. Moore$^{2}$,
Boris Pioline$^{3,4}$}

\bigskip

\centerline{$^{1}${\it Department of Theoretical Physics, Tata
Institute of Fundamental Research,}}

\centerline{\it  Homi Bhabha Road, Mumbai 400005, India}

\medskip

\centerline{$^{2}${\it Department of Physics, Rutgers University}}
\centerline{\it Piscataway, NJ 08854-8019, USA}

\medskip

\centerline{$^{3}${\it LPTHE, Universit\'es Paris 6 et 7, 4 place
Jussieu,}} \centerline{\it  75252 Paris cedex 05, France}

\medskip

\centerline{$^{4}$ \it LPTENS, D\'epartement de Physique de l'ENS,
24 rue Lhomond,}

\centerline{\it 75231 Paris cedex 05, France}

 \vskip.1in \vskip.1in \centerline{\bf Abstract}
 \medskip
\noindent We examine the recently proposed relations between black
hole entropy and the topological string in the context of type
II/heterotic string dual models. We consider the degeneracies of
perturbative heterotic BPS states.
In several examples
with $\CN=4$ and $\CN=2$ supersymmetry, we show that the
macroscopic degeneracy of small black holes agrees to all orders
with the microscopic degeneracy,
but misses non-perturbative corrections which are computable
in the heterotic dual.  
Using these examples we refine the previous proposals and comment
on their domain of validity as well as on  the relevance of helicity
supertraces.

\Date{February 16, 2005; Revised April 21, 2005; May 5, 2005; August 2, 2005}

%\draftmode

\def\IZ{{\bf Z}}
\def\IC{{\bf C}}
\def\IR{{\bf R}}
\def\mod{{\rm mod}}

\newsec{Introduction}

One of the distinct successes of string theory is that, in some
examples,  it gives an account of black hole entropy in terms of
statistical counting of microstates \refs{\StromingerSH,
\DamourKW}. One particularly rich set of examples are the BPS
black holes associated with D-branes wrapped on Calabi-Yau
manifolds in the type II string. In this case, the black hole
solutions exhibit fixed-point attractor behavior near the horizon
\refs{\FerraraIH, \StromingerKF}. Lopes Cardoso, de Wit, and
Mohaupt \refs{\LopesCardosoWT, \LopesCardosoCV,
\LopesCardosoXN,\LopesCardosoUR,\MohauptMJ} derived the generalized attractor
equations in the presence of higher derivative F-type terms and
obtained a formula for the Bekenstein-Hawking-Wald entropy of a
black hole \refs{\BekensteinUR, \BekensteinAX, \HawkingSW,
\WaldNT, \IyerYS}. Recently, Ooguri, Strominger, and Vafa 
proposed that the thermodynamical ensemble implicit in the above
entropy is a ``mixed" ensemble where magnetic charges are treated
micro-canonically while the electric ones are treated canonically  \OoguriZV.
This implies the following very elegant relation between the
topological string associated to the Calabi-Yau manifold $\CX$
\refs{\WittenXJ,\WittenZZ,\BershadskyCX, \BershadskyTA,
\AntoniadisZE} and the exact degeneracies of BPS states in the
theory.

In the Type-IIA string, the relevant BPS states arise from
wrapping D-branes on the various even cycles of the Calabi-Yau and
hence carry electric and magnetic charges denoted by a vector
$\gamma \in H^{\rm even}(\CX,\IZ)$. Upon choosing a symplectic
splitting, one can define the (magnetic, electric) charge
components of $\gamma$ as $(p^I,q_I), I = 0,1, \ldots, h^{1,1} (
\CX) $.  Moreover, on the moduli space of complexified K\"ahler
structures on $\CX$, one has a set of ``special coordinates'' $\{
X^I \}$. Let $F_{\rm top}$ denote the holomorphic topological string
partition function in these coordinates and define $\psi_p(\phi) =
e^{   F_{\rm top}(p+ i \phi)}$. Then, \OoguriZV\ proposes
\eqn\osvii{ \Omega(p,q) = \int d\phi\ \vert \psi_p  \vert^2
 e^{\pi q \cdot \phi },
}
where $\Omega(p, q)$ denotes the number or perhaps the ``index''
of BPS states of charges $(p^I, q_I)$. A weaker form of the conjecture 
requires that this equation holds
only to all orders in an asymptotic expansion in inverse charges
\OoguriZV. Equation\ \osvii has
in turn been reformulated 
in terms of a pure density matrix  in the  geometric
quantization of $H^{\rm 3}(\tilde{\CX},\IR)$ of the mirror
Calabi-Yau $\tilde{\CX}$ in the Type-IIB description \VerlindeCK.

While elegant, these formulae are somewhat imprecise.
The measure   $d\phi$ and the contour of integration in
the integral have  not been clearly specified, and the
precise choice of definition of the
microcanonical degeneracies $\Omega(p,q)$ has remained an issue.
In this note we report on some
attempts  to refine the proposal \osvii, and to test its accuracy
in explicit examples. A second paper in preparation will give further details
\DDMP.

In \DabholkarYR\  it was pointed out that type IIA/heterotic
duality offers a useful way to test \osvii, and this test was
initiated for the standard example of the $\CN=4$ duality between
 the heterotic string on $ T^6$ and the type IIA string on  $ K3 \times
T^2$. The main point is that there is an interesting class of BPS
states, the perturbative heterotic BPS states, (also known as
Dabholkar-Harvey states, or DH states, for short
\refs{\DabholkarJT, \DabholkarYF}), for which the exact
degeneracies are known or can be deduced using available string
technology. Moreover, much is known about the topological string
partition function in these examples. The present paper develops
further the use of type II/heterotic duality as a testing ground
for  \osvii.

The black holes corresponding to the DH states are mildly singular
in the leading supergravity approximation. The geometry has a null
singularity that coincides with the horizon and hence the
classical area of these black holes vanishes \refs{\SenEB,
\DabholkarNC}. Effects of higher derivative terms in the string
effective action are expected to modify the geometry \refs{\SenIN,
\SenIS}. Indeed, for  a subclass of higher derivative terms that
are determined by the topological string amplitudes,  the
corrected black hole solution can be determined  using the
generalized attractor equations \refs{\LopesCardosoWT,
\LopesCardosoCV, \LopesCardosoXN, \LopesCardosoUR,
\LopesCardosoQM}. The corrected solution has a smooth horizon with
string scale area in the heterotic string metric
\refs{\DabholkarYR, \DabholkarDQ,\SenDP, \HubenyJI,\BakMT}. We refer to
these black holes as `small' black holes\foot{The heterotic string
coupling becomes very small at the horizon and as a result the
horizon area is {\sl large} in the duality invariant Einstein
metric.} to distinguish them from the `large' black holes that
have large classical area.

Since small black holes have zero classical area, it is not {\it a
priori} obvious that the formula \osvii\ should  apply. However,
as noted above, the quantum corrected solution has a nonzero
horizon area. Combined with the successful determination of
degeneracies to all orders in $1/Q^2$ that we will find in
$\S{4}$, this gives strong evidence that there is nothing
particularly pathological about these black holes.
Nevertheless it should be borne in mind that the
$\alpha'$ corrections to these geometries remain to be
understood better.

We now give a brief overview of the remainder of the paper.

In $\S{2}$, we show that in certain scaling limits of charges
one can evaluate  the integral \osvii\ in  a saddle point
approximation that neglects the contributions of worldsheet instantons
to $F_{\rm top}$. We explain that this gives the leading asymptotic
expansion to all orders in   $1/Q^2$ where $Q$ is the
graviphoton charge. We argue that the analysis can be reliably
carried out for large black holes at both strong and weak coupling.
Our analysis in fact suggests that the proposal
\osvii\ must be modified slightly. The modified version is given in
eq. $(2.31)$ below.  As a matter of fact, one
encounters serious difficulties in trying to make sense of the
integral in \osvii\ non-perturbatively. We comment on these
difficulties, which arise mainly from the contribution of
worldsheet instantons to the topological string amplitude,  in $\S{5}$.

In $\S{3}$, we compute exactly the
microscopic degeneracies of the DH states in a broad class of
heterotic orbifolds with $\CN=4$ and $\CN=2$ supersymmetry and
determine their asymptotics using the Rademacher formula reviewed
in the Appendix. We also compute the ``helicity supertraces''   \FerraraQD\
that count the number of BPS short representations that cannot
be combined into long representations.
 For $\CN=2$ compactifications this is
 the space-time counterpart of the ``new
supersymmetric index'' on the worldsheet \CecottiQH, as shown in
 \refs{\LercheZY,\LercheZZ,\HarveyFQ}. One of the advantages of the
states that we consider is that both the absolute number
and the helicity supertraces
are computable exactly.

In $\S{4}$ we examine several $\CN=2$ and $\CN=4$
models in detail. In the $\CN=4$ examples we find
remarkable agreement between the microscopic and
macroscopic degeneracies to all orders in $1/Q^2$.
This computation can be rigorously justified. In the 
$\CN=2$ examples of small black holes there turn out to be
important subtleties in implementing the formalism of 
\OoguriZV. These are discussed in $\S{2.4.4}$ and the 
conclusions.  
 
In $\S{5}$ we summarize our results, point out some open
questions, and try to draw some lessons from what we have found.

Finally, we remark that there is
a reciprocal version of the proposal \OoguriZV.
In terms of this ensemble the formula of
\LopesCardosoWT\ is translated to:
\eqn\osvi{ e^{\CF(p,\phi)} = \sum_q \Omega(p,q) e^{-\pi q\phi} }
Using our exact knowledge of degeneracies of DH states, one may try
to construct the black hole partition function on the right-hand side
and compare to the topological string amplitude. As we shall discuss in
detail in \DDMP, we find that the result bears a close resemblance to
{\it a sum over translates} of the topological string amplitude,
enforcing the expected periodicity  under imaginary shifts
$\phi \to \phi+ 2 i \ZZ$. This indicates that a theta series
based on the topological string amplitude may be the appropriate
monodromy-invariant object to represent the complete
black-hole partition function \PiolineVI.

\newsec{Macroscopic Degeneracies via Saddle Point Approximation }

\subsec{Large radius limit}

To determine the macroscopic degeneracies of small black holes,
let us begin by attempting to evaluate the integral in \osvii\ for
a general compact Calabi-Yau   manifold $\CX$. The interpretation of $\psi_p$ as a
wavefunction certainly suggests that \osvii\ should be an integral
over a vector space, and we expect it to be an integral over a
real subspace of $H^{\rm even}(\CX,\IC)$. We will find below that the
definition of the measure $d \phi$ is nontrivial, but for the moment
we take it to be the standard Euclidean measure.

Now, the holomorphic topological string partition function is only defined
as an asymptotic expansion in the topological string coupling near
some large radius limit (i.e. in a   neighborhood of a point of maximal unipotent
monodromy). In this limit we can write the holomorphic prepotential as
a perturbative part plus a part due to worldsheet instantons.
See for example \refs{\HosonoAV, \HosonoAX}. We will write
\eqn\largrad{
F_{\rm sugra} = F^{\rm pert} - {i W^2 \over 2^7 \pi}  F^{GW}
}
The perturbative part is
\eqn\alrpre{
F^{\rm pert}=
- {C_{abc}\over 6} {X^a X^b X^c\over
X^0}  - W^2 {c_{2a}\over 24 \cdot 64 }
{X^a\over X^0} .
}
Here $a,b,c=1, \ldots, h$, $h= h^{1,1}(\CX)$, label components with respect to
an integral basis of $H_2(\CX,\IZ)$ (which we also take to be a basis inside the
K\"ahler cone), while $C_{abc}$ are
the intersection numbers of dual 4-cycles of the Calabi-Yau.
$c_{2a}$ are the components of the second Chern class.   $W^2$ is the square of the
Weyl superfield described in \MohauptMJ.
The sum over worldsheet instantons is %
\eqn\gromovwitten{
F^{GW} = \sum_{h\geq 0,\beta\in H_2(\CX,\IZ)} N_{h,\beta} \, q^{\beta} \, \lambda^{2h-2}
}
Here $N_{h,\beta}$ are the (rational) Gromov-Witten invariants,
\eqn\defqbta{
q^{\beta} = e^{2\pi i \int_{\beta} (B+iJ) } = e^{2\pi i \beta_a {X^a\over X^0} }
}
where $\beta_a\geq 0 $ are components of $\beta$ with respect to an integral basis of
$H_2(\CX,\IZ)$, and  $\lambda^2 = \bigl({\pi \over 4 X^0}\bigr)^2 W^2$.
In the topological string literature a slightly different normalization of the
prepotential is used. The two are related by  $F_{\rm sugra} = - {i W^2\over 2^7 \pi} F_{\rm top}$.
The attractor equations set $W^2= 2^8$ so then $F_{\rm top} = {i \pi \over 2} F_{\rm sugra}$.

\subsec{Perturbative evaluation}

It is natural to expect that the ``perturbative part''
 should give a good approximation
to the integral, at least for large charges. We will discuss in
detail what is meant by ``large charges'' in $\S{ 2.4}$  below,
where we will justify the procedure of   looking for a consistent saddle point in \osvii\
where it is a good approximation to replace $F_{\rm sugra}$ by $F^{\rm pert}$
defined in \alrpre. Following \OoguriZV\ we must evaluate
\eqn\curlyff{ \CF^{\rm pert} := - \pi ~ \im F^{\rm pert}(p^I + i \phi^I, 256) }
for
$\phi^I$ real. We will set $p^0 =0$, as this leads 
to significant simplifications.
In this case  we find that the perturbative part of the free
energy is given by
\eqn\cfhptv{ \CF^{\rm pert} =     -  {\pi \over 6} {\hat C(p)\over \phi^0 } + {\pi \over 2}{ C_{bc}(p) \phi^b
\phi^c\over \phi^0 }
}
where
\eqn\dfsn{  C_{ab}(p)  = C_{abc} p^c, \quad C(p)  = C_{abc} p^a
p^b p^c , \quad \hat C(p) = C(p) + c_{2a}p^a.  }
  The perturbative part has a saddle point for
\eqn\sdlepoint{ \phi_*^a = -C^{ab}(p) q_b \phi^0_* \qquad , \qquad
\phi^0_* = \pm \sqrt{ {- \hat C(p) \over 6 \hat q_0   } }
}
where $C^{ab}(p)$ is the inverse matrix of $C_{ab}(p)$ and
\eqn\dfnqh{ \hat q_0 = q_0 - \half q_a C^{ab}(p)  q_b  }
is the natural   combinations of charges compatible with the unipotent monodromy.
(In particular, $\hat q_0$ is monodromy invariant.)
\foot{In general one should allow an extra quadratic polynomial in $X^I$ with
real coefficients in $F^{\rm pert}$, say $-\half A_{ab} X^a X^b - A_a X^a X^0 - A (X^0)^2$
where $A_{ab}, A_a, A$ are all {\it real}.
The only effect of these terms in the present context is a shift of the charges to
$  \tilde q_a := q_a + A_{ab} p^b+ A_a p^0 , \tilde q_0 = q_0 + A_a p^a + 2 A p^0$.
This will not affect our arguments so we drop these terms for simplicity.}
In evaluating the saddle-point integral we must bear in mind that $C_{ab}(p)$
has indefinite signature (for example, for
$p^a$ an ample divisor $C_{ab}(p)$ has signature $(1, h-1)$) and therefore
$\phi^a \phi^b/\phi^0$ should be pure imaginary. We will take $p^a$ such that
$\hat C(p)>0$, and thus we want $\hat q_0<0$.

The integral \osvii, retaining only \cfhptv, is
 Gaussian on $\phi^a$ and of Bessel type for
$\phi^0$. The precise choice of $\phi^0$ contour does not matter if we
only concern ourselves with the asymptotic expansion of the $\phi^0$ integral for
$\hat C(p) \vert \hat q_0 \vert \to + \infty$. The asymptotics can then be given in
terms of those of a Bessel function, the precise formula being:
\foot{If we want to get the actual Bessel function from the $\phi^0$ integral
then the appropriate contour to take is the circle described by
$1/\phi^0 = - \epsilon + i s$, $\epsilon>0$, $s\in \IR$. However,
we should not   discuss contours before the nonperturbative
completion of $\psi_p$ is specified. }
%The contour for the integral on $\phi^0$ could be taken to be
%parallel to the imaginary axis
%\foot{This Wick rotation is also required for the Wigner function
%interpretation advocated in \OoguriZV.}
%or one could take $1/\phi^0$ to be parallel to the imaginary axis.
%In either
%case
%and the integral    can be evaluated exactly as:
%
\eqn\ibessl{
\CN(p)\ 
\hat I_{\nu } \Biggl(2\pi \sqrt{- \hat C(p) \hat q_0\over  6} \Biggr) }
where $\hat I_\nu(z)$ is related to the Bessel function $I_\nu(z)$ as
in equation $(A.3)$ of the appendix and
\eqn\nugenrl{
\nu = \half(n_v+1) .
}
Here $n_v$ is the
rank of the total 4-dimensional gauge group, so $n_v= h+1$.
 The Bessel function grows
exponentially, for large ${\rm Re}(z)$
(see $(A.6)$ )  so that the leading asymptotics of
\ibessl\ agrees with the standard formula from  \MaldacenaDE\
evaluated in the same limit.
The factor $\CN(p)$ is given by
\eqn\ennofp{
\CN(p) = \pm \half   \sqrt{1 \over \vert \det C_{ab}(p)\vert } \Bigl({\hat C(p)  \over 6} \Bigr)^{\nu}
}
and only depends on the magnetic charges $p^a$ and not on the electric
charges $q_a$.

\subsec{Modifications for small black holes}

By definition, a small black hole is a BPS state such that $C(p)=0$ but $\hat C(p) \not=0$.
In this case, while the horizon is singular and of zero area in the classical supergravity,
it is expected that quantum corrections will smooth out the singularity leading to a
legitimate black hole. For such charges, some of the manipulations in the previous
section are not valid and must be modified as follows.

We are particularly interested in the case when $\CX$ is a $K3$
fibration over $\IP^1$ admitting a heterotic dual. Moreover, we
are interested in charges corresponding, on the heterotic side, to
DH states. As we will see, we cannot simply plug into \ibessl. Nevertheless, a
similar computation applies. If $\CX$ is K3-fibered then we can
  divide up the special coordinates so that $X^1/X^0$ is the
volume of the base and $X^a/X^0$, $a=2, \dots n_v -1$ are
associated with the
(invariant part of the) Picard lattice of the fiber.
The charges of heterotic DH states have   $p^0=0$, $ p^a=0, a=2,\dots, h$, and
$q_1=0$, with $p^1q_0 \not=0$ and $q_a\not=0$ for $a=2,\dots, h$.
In this case $C_{ab}(p)$ is of the form
\eqn\dabmatrx{
p^1\pmatrix{0 & 0 \cr 0 & \tilde C_{a'b'} \cr}
}
where $\tilde C_{a'b'}$ is the intersection form of the (invariant part of the)  Picard lattice of the
fiber. Note that now $C_{ab}(p)$ is not invertible. The $\phi^1$ dependence disappears from the
integrand and one must  make a discrete identification on $\theta
= \phi^1/\phi^0$. One thereby finds that \osvii\  gives
\eqn\iresult{ \CN(p) \ \hat I_{\nu}\bigl( 4\pi \sqrt{\vert p^1 q_0 - \half
q_{a'}\tilde{ C}^{a'b'} q_{b'} \vert }\bigr) }
where
\eqn\iresulti{ \nu = \half (n_v+2) }
 and $\CN(p)$ is a $p$-dependent prefactor.

Note that the argument of the Bessel function \ibessl\ nicely
reduces to that of \iresult. For DH states, $C(p)=0$, reflecting
the fact that the classical area of the corresponding black holes
 is zero, and
the nonzero entropy is provided by the quantum correction
$ c_{2a} p^a$. The change in index of the Bessel function results
from an enhanced volume factor $\sqrt{\phi^0}$ arising from the
zero mode of $C_{ab}(p)$.

Comparison with the exact results on DH degeneracies below shows
that there is a nontrivial question of how to normalize the
measure $d\phi$ (or  the wavefunction $\psi_p $). In particular
 the $p$-dependent prefactors $\CN(p)$ in \ibessl\iresult\ are not compatible
with exact results.  An important point revealed by the case of
small black holes is that the wavefunction
$\psi_p$ is  in fact not  normalizable, at least, not
 in the conventional sense. We will
return to this in the discussion section at the end.

\subsec{Justification of the saddle point evaluation}

\subsubsec{\ \it Large charge limits}

In this section we show that the
perturbative evaluation of the integral
performed above is valid provided we consider an appropriate
scaling limit of large charges.

Let us begin by considering  a rather general scaling limit of charges
\eqn\chargscale{
\eqalign{
\hat q_0 & \to s^{x}  \hat q_0 \cr
p^a & \to s^y p^a  \cr}
}
where $s\to \infty$. Here $x,y\geq 0$ and $p^a$ defines a
vector in the K\"ahler cone.  This scaling will result in a scaling
\eqn\sdpv{
\phi^0_* \to s^z  \phi^0_* + o(s^z)
}
for the saddle-point value   $\phi^0_*$. Here $o(s^z)$ means
terms growing strictly more slowly than $s^z$. (For example, 
from  the saddle-point equation $(2.20)$ below  
$z=(3y-x)/2$.) 
Now, there are three  criteria we might wish to impose in order to be
able to evaluate the integral \osvii\ reliably in the saddle point
approximation:

\item{1.} {\it Neglect of worldsheet instantons}.
We expect the worldsheet instanton series to be small if $\im {X^a\over X^0} \gg 1$.
In the saddle point approximation this means we require
\eqn\spone{
- {p^a \over \phi^0_*} \gg 1 .
}
for all $a$. We fix the overall sign by choosing $p^a>0$ and hence
$\phi^0_*<0$. Having all $p^a>0$ means the divisor wrapped by the
D4-brane is very ample. The above criterion requires $y> z$.

\item{2.} {\it Weak coupling in the expansion in $\lambda$}. A natural condition to require
 is that the topological string is weakly coupled.
Physically, this is the requirement that the expansion of the supergravity effective
action in powers of the graviphoton fieldstrength is not strongly coupled.
Using the attractor value $W^2 = 2^8$ this means $\lambda = - 4\pi i /\phi^0_*$ is small.
Hence we require $z>0$ for weak topological string coupling.

\item{3.} {\it Saddle-point equations}. We insist that $\phi^0_*$ satisfy the saddle
point equations for the relevant approximation to $\CF$.
 In the case of a weakly coupled
topological string we must add the term
\eqn\deltaeff{
\Delta \CF =  { \zeta(3) \chi(\CX) \over (4\pi)^2} (\phi^0)^2 := {\pi\over 2} \xi (\phi^0)^2
}
to \cfhptv.  Thus the explicit equations are
\eqn\speqs{
\eqalign{
 {\hat C(p)\over 6} + \hat q_0 (\phi^0)^2 + \xi (\phi^0)^3 =0 \qquad& \qquad {\rm weak\quad coupling} \cr
   {\hat C(p)\over 6} + \hat q_0 (\phi^0)^2  =0 \qquad\qquad\qquad & \qquad {\rm strong\quad coupling\quad and/or }\quad \chi(\CX)=0 \cr}
}
The full justification of the second line of \speqs\ is given in $\S{2.4.2}$. 

There are  two important subtleties in imposing the condition \spone.
First the
Gromov-Witten series \gromovwitten\ includes the contribution of {\it pointlike instantons}
with $\beta=0$, and the criterion \spone\ does not lead to suppression of these terms,
which must therefore be considered separately. Second there are further subtleties for
small black holes discussed in   $\S{2.4.4}$  below. Sections $\S{2.4.2}$ and 
$\S{2.4.3}$ concern large black holes. Readers only interested in small 
black holes should skip to   $\S{2.4.4}$.

While weak coupling is a natural condition to impose, we will argue
that it is not always necessary to do so, and of course one wants 
to understand both weak and strong coupling limits. In some cases, 
such as the small $\CN=4$ black holes, the computation of the 
macroscopic degeneracy can be fully justified at weak coupling
(and turns out to be the same as at strong coupling).

\subsubsec{\ \it Strong topological string coupling}

There are certain charge limits of great interest in which 
one must work at strong topological string coupling. 
For example, in order to
compare asymptotic degeneracies in the dual CFT description  of
\MaldacenaDE\ one requires that the level number be much larger than
the central charge, and hence
\eqn\sptwo{
\vert \hat q_0 \vert \gg \hat C(p)
}
(Validity of the supergravity approximation leads to a similar, but
less restrictive criterion $\vert \hat q_0^3\vert \gg C(p) $
\MaldacenaDE.) Equation \sptwo\  imposes the condition $x > 3y$ for
large black holes. It is easy
to see that in either case, the condition \sptwo\ is incompatible
with \spone\speqs\ and weak coupling. This motivates us to take a
closer look at strong topological string coupling.

In this section we consider
the limit of charges \sptwo, and we will argue that it suffices to
use the approximation \cfhptv\ in this case. Thus, from \sdlepoint\
  the topological string coupling
$\lambda= -4\pi i/\phi^0_*$ is large, and therefore the
topological string is strongly coupled.

In order to justify our procedure we separate the pointlike
instantons from those with nonzero area by writing
\eqn\spfour{
F^{GW} = F^{GW}_{\beta=0} + F^{GW}_{\beta \not=0}
}

First, let us consider $F^{GW}_{\beta \not=0}$.
The  worldsheet instanton corrections with $\beta \not=0$ are
formally suppressed by
\eqn\spthree{
\CO\biggl(e^{-2\pi p^a \beta_a\sqrt{6 \vert \hat q_0 \vert \over \hat C(p)} }\biggr)
}
where $\beta_a \geq 0 $. Hence one may
formally neglect the $\beta\not=0$ terms in  $F^{GW}$ up to exponentially
small corrections. One should be careful at this point.
Since the nonperturbative completion of the topological
string is not known we must make an assumption.
We will simply assume that $F^{GW}_{\beta \not=0}$ has a nonperturbative
completion so that the formal suppression \spthree\ is valid,
even though $\lambda \to \infty$. The justification of this
assumption awaits a nonperturbative definition of the topological
string. Nevertheless, let us note that this is a very reasonable
assumption. The   key point is that although the
topological string coupling $\lambda$ goes to infinity,
{\it the K\"ahler classes also go to infinity}.
\foot{This remark also resolves the following puzzle: If $\lambda$ is
large one might expect the genus one term to dominate over the genus
zero term. In fact, they are both of the same order, as is evident
from \cfhptv. }
The reason is that
at the saddle point, $\im t^a = p^a \vert \lambda\vert $.
Thus the contribution $\lambda^{2h-2} q^\beta $ for $h>1$
behaves like $\lambda^{2h-2} e^{-\kappa  \lambda}$ where
$\kappa $ is a positive constant. It therefore decays exponentially
fast, even at strong coupling. More precisely, the contribution is
\eqn\hightrms{
N_{h, \beta} \biggl( {\vert \hat q_0 \vert \over \hat C(p)} \biggr)^{h-1}
e^{-2\pi p^a \beta_a\sqrt{6 \vert \hat q_0 \vert \over \hat C(p)} }
}
and in the limit \sptwo\ this vanishes rapidly. 

The above hypothesis can also be partially justified using the 
infinite product representation of $\exp F_{\rm top}$ 
implied by the work of Gopakumar and Vafa \GopakumarJQ. 
The infinite product may be split into three factors 
involving the   BPS (a.k.a. Gopakumar-Vafa) invariants $n^{(h)}_\beta$ of spins $h=0$, 
$h=1$ and $h>1$. The infinite products involving spin $h=0$ and spin $h=1$ 
BPS invariants can be shown to be convergent in appropriate 
domains, and they indeed satisfy our hypothesis. Unfortunately the infinite products 
involving spin $h>1$ BPS invariants are in general not convergent. 
(The problem is that the maximal spin $h_*(\beta)$ for which  $n^{(h)}_\beta$
is nonzero grows too rapidly with $\beta$.) Thus, in general,  we cannot use the 
infinite product representation to give a nonperturbative definition. 
However, if $n^{(h)}_\beta=0 $ for $h>1$ then our hypothesis is rigorously justified.

Now we must turn to the effects of the pointlike instantons
contributing to $F^{GW}_{\beta=0}$.
The results of \GopakumarJQ\ lead to a nonperturbative completion of
$F^{GW}_{\beta=0}$. We have
\foot{This identity  is not stated correctly in the topological string theory literature,
which omits the second and third terms on the left-hand side.}
\eqn\gvidnet{
    n^0_0 \Biggl[ f(\la)  + {1\over 12} \log {\la\over 2\pi i } - K\Biggr]   \sim \sum_{h}  N_{h,0} \la^{2h-2}
}
where
\eqn\hzergw{
\sum_{h}  N_{h,0} \la^{2h-2}=   -\half \chi(\CX) \Biggl[\lambda^{-2} \zeta(3)- \sum_{n=0}^\infty \lambda^{2n+2} {\vert B_{2n+4} \vert \over (2n+4)!}
 {(2n+3)  \over (2n+2)  }    B_{2n+2}  \Biggr]
}
for $\lambda\to 0$. Here $n^0_0 = - \half \chi(\CX)$,  $K=  -{1\over 24} - {\zeta'(2)\over 2\pi^2} + {\gamma_E\over 12}$
is a constant, and
\eqn\mcham{
f(\lambda):= \sum_{d=1}^\infty {1\over d} (2\sin {d\la \over 2})^{-2}=  \log \prod_{k=1}^\infty (1-e^{i \la k})^k .
}
(the second identity holds for $\im \lambda>0$). The important point is that the left-hand side of \mcham\ is a well-defined function of
$\lambda$, so long as $\lambda\notin \IR$, and therefore defines a nonperturbative
completion of $F^{\rm GW}_{\beta=0}$. Using the infinite-product (McMahon) 
formula for $f(\la)$ we have
\eqn\nsvnt{
e^{\CF^{\rm GW}_{\beta=0} } = \bigl(-{\phi^0\over 2}\bigr)^{\chi/12} e^{K \chi}
\Bigl( {\prod_{k\geq 1} (1-e^{4\pi k \over \phi^0})^k} \Bigr)^{-\chi}
}
for $\phi^0<0$. Now, for $\phi^0 = - \sqrt{\hat C/6\hat q_0}$ negative and small, the
infinite product is $1 + \CO(e^{-4\pi \sqrt{6\vert \hat q_0\vert/\hat C}}) $.

The factor $\bigl(-{\phi^0\over 2}\bigr)^{\chi/12} $ in \nsvnt\
will spoil the remarkable agreement between \osvii\ and certain states in
$\CN=2$ models with $\chi\not=0$, as described below.
Therefore, to preserve this success we modify by hand the topological string wavefunction
\eqn\modtopstr{
\Psi_{top}\to \tilde \Psi_{top} := \lambda^{\chi/24} e^{  F_{\rm top}}
}
so that
\eqn\modfed{
\tilde \psi_p(\phi):= \bigl(-{\phi^0\over 2}\bigr)^{-\chi/24} e^{F_{\rm top}(p+i \phi) }
}
and we propose a modification of the conjecture \osvii:
\eqn\modosv{
\Omega(p,q) = \CM(p) \int d \phi \vert \tilde \psi_p(\phi)\vert^2 e^{\pi q \phi}
}
where $\CM(p)$ depends on $p$ but not on $q$. This normalization
factor is unavoidable; the $p$-dependent factor arising from the
integrations, such as \ennofp, in general does not agree with the
$p$-dependent prefactor of the asymptotic expansion of the
microscopic index.

To summarize, the integral in \modosv\ may be defined as an asymptotic expansion in
charges in the scaling limit \sptwo. The value of the integral is
\eqn\summrze{
\CN(p)\ 
\hat I_{\nu } \Biggl(2\pi \sqrt{\hat C(p) \vert \hat q_0\vert \over  6} \Biggr)
\cdot \Biggl( 1 + \CO(e^{-\kappa(p)\sqrt{\vert \hat q_0\vert}} ) \Biggr)
}
where $\CN(p), \kappa(p)$ are $p$-dependent constants.

The modification \modfed\ is very similar to an extra factor $\lambda^{\chi/24 -1}$
which is included in the {\it nonholomorphic} topological string wavefunction. See
 \refs{\BershadskyCX,\BershadskyTA,\VerlindeCK}.  We expect that
taking proper account of measure factors in the definition of the wavefunction as a
half-density  will lead to a more satisfactory
justification of our modification \modtopstr.

\subsubsec{\ \it Weak  topological string coupling}

Now let us consider the situation for weak coupling. This can be achieved with
a limit of charges  with
\eqn\weakcoup{
y<x < 3y
}
If $\chi(\CX)\not=0$ then the saddle point equation in \speqs\ has three roots.
The discriminant is
$$
{\hat C \over 12 \xi}\biggl( {\hat C \over 12 \xi} + 2 \bigl({\hat q_0 \over 3 \xi} \bigr)^3 \biggr)
$$
and hence if  $y<x$ there are three real roots of \speqs. One root $\phi^0_* \sim - \hat q_0/\xi + \cdots$ is
inconsistent with large K\"ahler classes. The other two roots are
\eqn\exproot{ \phi^0_*  = \pm \sqrt{\hat C(p) \over 6 \vert \hat
q_0\vert} \Biggl( 1 \mp \half \xi   \sqrt{\hat C(p)\over 6 \vert \hat
q_0\vert^3} + \cdots\Biggr) }
and as discussed earlier we choose the negative root. The
saddlepoint evaluation of the integral is proportional to
\eqn\weksddlept{
(\det C_{ab}(p))^{-1/2} \int d \phi^0  (\phi^0)^{h/2} \exp\biggl[
-{\pi \hat C \over 6 \phi^0} + \pi \hat q_0 \phi^0 + {\pi \over 2} \xi (\phi^0)^2 +
\sum_{h=2}^\infty N_{h,0} \bigl({4\pi i \over \phi^0}\bigr)^{2h-2} \biggr]
}
evaluated in an expansion around \exproot. (If we use the modified version \modosv\
then we must replace $(\phi^0)^{h/2} \to (\phi^0)^{h/2-\chi/24} $ in \weksddlept.)
The asymptotics will no longer be governed by a Bessel function,
as in the strong couping regime. The leading correction to the entropy $2\pi \sqrt{\hat C \vert \hat q_0 \vert/6}$
is no longer of order $\log s$, as in \ibessl\ but rather grows like a 
positive power of $s$: 
\eqn\contribent{
S = 2\pi \sqrt{\hat C \vert \hat q_0 \vert\over 6} + {\zeta(3)\chi(\CX)\over 96\pi^2} {\hat C \over \vert \hat q_0 \vert} + \cdots
}
It is an interesting challenge to reproduce this from a microscopic 
computation \foot{A similar correction has been computed in \LopesCardosoUR, 
without taking
into account the contribution from the integration measure in
\osvii}.

Finally, for completeness we note that if $x<y$ then  (for $\chi\not= 0$) the roots are approximately
$\phi^0 \sim (-\hat C/6\xi)^{1/3}$ and the Kahler classes are small. This means that in this regime of charges
one must retain the full genus zero worldsheet instanton series.

\subsubsec{\ \it Additional subtleties for small black holes}

In the case of small black holes $C(p)=0$. Since the saddle point value of
$\im t^a = - p^a/\phi^0_*$, this implies that $C(\im t) =0$ and hence the saddle point
is necessarily at the boundary of the K\"ahler cone. In principle, one
must retain the full worldsheet instanton series (or rather, its analytic continuation,
should that exist.)

Remarkably, for $\CN=4$ compactifications this is not a problem. In this case
$F_{\rm top}$ is only a function of a single K\"ahler modulus, namely, $t^1$ in the
notation of  $\S{2.3}$. The reason is that the moduli space
factors as a double-coset of $SL(2,\IR)$ times a Grassmannian, and by
decoupling of vector and hypermultiplets, $F_{\rm top}$ must be constant
on the Grassmannian factor. Moreover, in these compactifications $\chi(\CX)=0$ and
hence the saddle-point values are:
\eqn\sdlsbh{
\phi^0_* = - \sqrt{4 p^1 \over   \vert \hat q_0 \vert} \qquad  \im t^1 = \half \sqrt{p^1 \vert \hat q_0 \vert}
}
Thus, whether or not the topological string coupling is strong ($\vert \hat q_0 \vert \gg p^1$) or
weak ($p^1\gg \vert \hat q_0 \vert $) the relevant K\"ahler class is large and the
Bessel asymptotics \iresult\ are justified.

The situation is rather different for $\CN=2$ compactifications. In this case
$F_{top}$ is in general a function of $t^1$ as well as $t^a$ for $a\geq 2$.
Thus the computation of section $\S{2.3}$ is {\it not} justified.
We stress that the problem is not  that the
topological string is strongly coupled. Indeed,  for $\chi=0$
examples such as the FHSV example discussed in  $\S{4.3}$ below,
the saddlepoint value \sdlsbh\ can be taken in the weak coupling regime
by taking $p^1\gg \vert \hat q_0 \vert $.
In fact, the difficulty appears to be with the formulation of the
integral \osvii\ itself
for the case of charges of small black holes.
Recall that we must evaluate
\eqn\curlyff{ \CF := - \pi ~ \im F(p^I + i \phi^I, 256) }
Since $X^a/X^0 = \phi^a/\phi^0$ is {\it real}, for $a>1$,  one must
evaluate the worldsheet instanton sum for real values $t^a = \phi^a/\phi^0$.
For some Calabi-Yau manifolds it is possible to analytically continue
the tree-level prepotential $F_0$ from  large radius 
to small values of $\im t^a$. However we may use  the
explicit results of \HarveyTS\HenningsonJZ, which express
 $F_1 \sim \log \Phi$, where $\Phi$ is an automorphic form for $SO(2,n;\IZ)$.
It appears that $\im t^a =0$
constitutes a natural boundary of the  automorphic form  $\Phi$. 
Thus the formalism of 
\OoguriZV\ becomes singular for these charges, even at weak topological string coupling.

Remarkably, if we ignore these subtleties, the formula \iresult\ turns out
to  match perfectly with the asymptotic expansions  of twisted sector
DH states, as we show below. For untwisted sector DH states the asymptotics
do not match with either the absolute degeneracies $\Omega_{abs}$ nor with 
the helicity supertrace $\Omega_2$, as discussed in Section 3.

\subsec{Holomorphic vs. non-holomorphic 
topological string partition functions}

The asymptotic expansion  of the integral \modosv\ differs from the
entropy predicted from
the attractor formalism, as modified in
\refs{\LopesCardosoWT, \LopesCardosoCV,
\LopesCardosoXN, \LopesCardosoUR,\MohauptMJ}.  The latter identifies
\eqn\spvalue{
S =
\biggl[ \CF - \phi^I {\p \CF \over \p \phi^I} \biggr]_{s.p.}.
}
This  is just  the leading semiclassical approximation to \osvii\ and does not
capture the subleading corrections given by the asymptotics of the Bessel
function. The same argument we have used to justify evaluating the
integral \modosv\ with $\CF^{\rm pert}$ can be applied to \spvalue.
After a suitable modification $\CF \to \tilde \CF = \CF - {\chi\over 12} \log \phi^0$
the entropy given by   \spvalue\ using the full nonperturbative prepotential $\tilde \CF$
is the same as that given by $\CF^{\rm pert}$, up to exponentially small corrections.
As we will see, this leads to predictions at variance with exact counting of
heterotic BPS states.

Several recent papers \refs{\SenDP, \CardosoXF, \SenPU, \deWitFZ } have addressed this
problem by taking into account the holomorphic anomaly in topological
string theory. In particular, in the paper
 \SenPU\   the microscopic and macroscopic degeneracies
for small black holes are shown to match in reduced rank $\CN=4$
models using a different ensemble than suggested by \osvii.
Roughly speaking, the idea is that one has instead
\eqn\spvalue{
S =
\biggl[ \CF_{\rm eff}  - \phi^I {\p \CF_{\rm eff} \over \p \phi^I} \biggr]_{s.p.}.
}
where $\CF_{\rm eff} $ is a non-Wilsonian, non-holomorphic effective action.
On the other hand, it is clear
from the discussion in \VerlindeCK\ that one should use the {\it holomorphic}
prepotential in \osvii\modosv.
These two approaches are not necessarily incompatible. The nonholomorphic
effective action is obtained from the holomorphic Wilsonian effective action
by integrating out massless modes. In a similar way  $\CF_{\rm eff}$
might in fact be defined by carrying out the integral \osvii\modosv.

\newsec{Microscopic Degeneracies of Heterotic DH States}

Let us now determine   the microscopic degeneracies of the DH
states using the heterotic dual. For concreteness, we will focus
here on bosonic orbifolds of the heterotic string on $T^6$. (Using
the elliptic genus it should be possible to extend the  results in this section
to a wider class of models.) We will denote the orbifold group by
$\Gamma$. There is an embedding $R: \Gamma \to O(22)\times O(6)$.
The orbifold group also acts by shifts so that the action on
momentum vectors is
\eqn\momact{ g \vert P \rangle = e^{2\pi i \delta(g)\cdot P} \vert
R(g) P \rangle.
 }
In $\IR^{22,6}$, with metric $Diag(-1^{22},+1^{6})$  we can
diagonalize the action of $R(g)$ with rotation angles $2\pi
\theta_j(g)$, $j=1,\dots, 11$ on the leftmoving space and $2\pi
\tilde \theta_j(g)$, $j=1,2,3$ on the rightmoving space. The
moduli are the boosts in $O(22,6)$ commuting with the image
$R(\Gamma)$. We consider embeddings $\Lambda\subset \IR^{22,6}$ of
$II^{22,6}$. We let $\Lambda(g)$ denote the sublattice of vectors
fixed by the group element $g$. Of course, there will be
constraints from level matching and anomaly cancellation. We
assume that those constraints are satisfied. This still leaves a
large class of possibilities.

$\CN=1$ spacetime supersymmetry requires that $\sum_i \tilde
\theta_i(g) = 0 ~\mod~1$ for all $g$. $\CN=2$ spacetime
supersymmetry requires that   $\tilde \theta_3(g)=0$ for all $g$.
In this case we let $\tilde \theta(g) := \tilde
\theta_1(g)=-\tilde \theta_2(g)$. $\CN=4$ spacetime supersymmetry
requires $\tilde \theta_i(g)=0$ for all $i,g$.

The orbifold model will have a gauge symmetry. The currents in the
Cartan subalgebra of the gauge symmetry (which is generically
abelian) is spanned  by $k$ pairs of left-moving bosons which are
fixed for all $g\in \Gamma$, i.e. we suppose $\theta_i(g)=0$ for
all $g$ for $i=1, \dots, k$. \foot{For brevity we restrict   some
generality. It is possible to have $\theta_i(g)= \half$ allowing
an odd number of twisted bosons. The formulae below are easily
modified to accommodate this case.} There is a subspace   $\CQ
\subset \IR^{22,6}$  fixed by all group elements. It is of
signature $(2k,6)$ for $\CN=4$ compactifications and
$(2k,2)$ for $\CN=2$ compactifications, respectively. 
The vector-multiplet moduli
come from the $SO(2k,6)$ (resp. $SO(2k,2)$) rotations in this plane.
The number of $U(1)$ vector fields is $n_v=2k+6$ in the $\CN=4$
compactifications and $n_v = 2k+2$ in the $\CN=2$
compactifications. The lattice of electric charges (in the
untwisted sector) for the gauge symmetry is the orthogonal
projection (in the metric $(-1^{22},+1^6)$) of $\Lambda$ into the
plane $\CQ$.   Denote the charge lattice in the untwisted sector
by $M_0$ and let $Q_{el}: \Lambda \to M_0$ be the orthogonal
projection. States in the untwisted sector are naturally labelled
by $P \in II^{22,6}$ but we only want to compute degeneracies at a
fixed charge vector $Q\in M_0$.

Let us now compute the degeneracies of the DH states. In the
untwisted sector DH states are all contained in the subspace of
the 1-string Hilbert space of the form
\eqn\thens{ \CH_{osc,L} \otimes \CH_{mom} \otimes \tilde \CH_{gnd}
}
satisfying $L_0 = \tilde L_0$. Here the three factors are
leftmoving oscillators, momentum eigenstates, and rightmoving
groundstates. One important subtlety which arises for $\CN=2$
compactifications
 is that, in general,  even in this subspace the DH states span a
proper subspace. The projection to the BPS states   depends only
on the momentum $P$ of the state and implements the BPS condition
$M^2 = (Q_{el})^2$. Let $\Pi_{bps}(P)$ be $1$ if this
condition is satisfied, and zero otherwise. For some vectors $P$
we have $\Pi_{bps}(P)=1$ throughout the entire moduli space.
However there can also be ``chaotic BPS states'' for which
$\Pi_{bps}(P)=0$ generically but, on a subspace of hypermultiplet
moduli space, jumps to one \HarveyFQ.

The space of BPS states is graded by the electric charge lattice
$M_{el}$ (in general $M_0$ is a proper sublattice) and we denote
by $\CH_{BPS}(Q)$ the subspace with charge $Q$. We will be
interested in several measures of the degeneracies of states. The
absolute number is $\Omega_{abs}(Q):= \dim \CH_{BPS}(Q)$.
   Because of the chaotic BPS states this is not a constant function on
moduli space. Examples show that a more appropriate quantity for
comparing to \osvii\ are the helicity supertraces. These
are defined by
\foot{These supertraces generalize the 
``vectors minus hypers'' index used in \HarveyFQ. See
\KiritsisHJ\ appendix G for a nice discussion of helicity
supertraces.}
\eqn\beetwo{ \Omega_{n}(Q) := {1\over 2^n} \bigl(y{\p \over \p
y}\bigr)^n \vert_{y=+1}   {\Tr}_{\CH_{BPS}(Q)} (-1)^{2J_3}
y^{2J_3} }
where $J_3$ is a generator of the massive little group in 4 dimensions.
For $\CN=2$ compactifications the first nonvanishing supertrace is
 $\Omega_2(Q)$ and this appears to be the correct quantity to use when
comparing with the integral \osvii. Only BPS states contribute to
$\Omega_2(Q)$.
%%%%
%%%%
For $\CN=4$ compactifications
the first nonvanishing supertrace is $\Omega_4(Q)$. This only
receives contributions from $\half$-BPS states. For $\Omega_6(Q)$ both
$\half$- and ${1\over 4}$-BPS states contribute. Examples suggest
that $\Omega_4(Q)$ is the appropriate index to use for $\half$-BPS
states. Clearly, a different index must be chosen for 1/4-BPS states, if
Eq. \osvii\  is to continue to hold for them as well. $\Omega_6(Q)$
is then the only candidate in this case.
%%%%
%%%%

The evaluation of the partition function in the BPS subspace of
\thens\ is largely standard. Care must be exercised in the
evaluation of the momentum sum since we are only interested in the
degeneracies of the BPS states at a fixed $Q\in M_{el}$.  In the
untwisted sector we should write the momentum contribution as:
\eqn\lattsum{
 \sum_{P\in \Lambda(g)} q^{\half P_L^2} \bar q^{\half P_R^2}
e^{2\pi i \delta(g)P} \Pi_{bps}(P) = \sum_{Q\in M_0} q^{\half
Q_L^2} \bar q^{\half Q_R^2} \CF_{g,Q}(q) }
where
\eqn\extraeff{ \CF_{g,Q}(q) = \sum_{P\in \Lambda(g), Q_{el}(P)=Q}
q^{\half (P_L^2- Q_L^2)} e^{2\pi i \delta(g)P} \Pi_{bps}(P) }
Note we have used the BPS condition $P_R^2 = Q_R^2$, and due to
this condition we can write $P_L^2-Q_L^2 = P^2-Q^2$. The function
\extraeff\ is actually very simple in many important cases. For
example if $\Lambda(g) \subset M_0$, which is typical if the fixed
space under the group element $g$ coincides with $\CQ$ then we
simply have $\CF_{g,Q}(q) = e^{2\pi i \delta(g)\cdot Q} $. For
this reason it is useful to distinguish between ``{\it minimal
twists}'', which leave only the subspace $\CQ$ invariant (i.e. $0
< \theta_j(g)<1$ for $j>k$) and nonminimal twists. For nonminimal
twists the kernel of $Q_{el}$ will be nontrivial and
$\CF_{g,Q}(q)$ will be a theta function.

Putting all this together the degeneracies of untwisted sector BPS
states are given by
\eqn\bsnttn{
 \Omega_{n}(Q) =  e^{4\pi Q_R^2} \int d\tau_1\ q^{\half Q_L^2} \bar
q^{\half Q_R^2} \CZ_n }
where
\eqn\andthcz{ \CZ_n = {1\over  \vert \Gamma\vert } \sum_{g\in \Gamma} {1\over
\eta^{2+2k} }   \Biggl[ \prod_{j=1}^{11-k} (-2\sin \pi
\theta_j(g)) {\eta \over \vt{\half}{\half + \theta_j(g)}{} }
\Biggr] w_n(g) \CF_{g,Q}(q) }
and $w_n(g)$ is given by
\eqn\omegacases{ w_n(g) = \cases{ 16 \cos \pi \tilde \theta_1(g)
\cos \pi \tilde \theta_2(g)\cos \pi \tilde \theta_3(g)  & $n=abs$
\cr 2 (\sin \pi \tilde \theta(g))^2 & $n=2$ \cr
 {3\over 2} & $n=4$\cr
{15\over 8}(2-E_2(\tau)) & $n=6$\cr} }

The formula \andthcz\ is exact.  Quite generally, the partition
functions are negative weight modular forms and the degeneracies
are given by their Fourier coefficients. There  is a general
formula - the Rademacher expansion - for the coefficients of such
modular forms which is exact and yet  summarizes
beautifully the asymptotic behavior of these coefficients. It
expresses these coefficients as an infinite sum of $I$-Bessel
functions and thus is very well suited to comparison with the
integral expression \iresult. The Rademacher expansion is
summarized in the appendix.

Using the Rademacher expansion, the leading asymptotics for the
degeneracies of DH states from the minimal twists is ($n\not=6$
here):
\eqn\asympts{
   {1\over  4\vert \Gamma\vert } \sum_{g\in \Gamma, minimal }' w_n(g) h(g)
\prod_{j=1}^{11-k}(-2\sin \pi \theta_j(g)) \vert
\Delta_g\vert^{k+2}  \hat I_{k+2 }(4\pi \sqrt{\vert
\Delta_g\vert \half Q^2  }) }
where
\eqn\hgee{ h(g) = \cases{ (-1)^{(12-k)/2} \sin\bigl(2\pi \delta(g)
Q + \pi \sum_j \theta_j(g)\bigr)  & $k$ even \cr (-1)^{(11-k)/2}
\cos\bigl(2\pi \delta(g) Q + \pi \sum_j \theta_j(g)\bigr)  & $k$
odd \cr} }
and
\eqn\deltagee{ \Delta_g := -1 + \half \sum_{j=1}^{11-k}
\theta_j(g)(1-\theta_j(g)) , \qquad\qquad 0< \theta_j(g) < 1 }
is the oscillator ground state energy in the sector twisted by
$g$. The prime on the sum indicates we only get contributions from
$g$ such that $\Delta_g <0$. For nonminimal twists there will be
similar contributions as described above. In particular the index
on the Bessel function will be the same, but \deltagee\ receives
an extra nonnegative contribution from the shift $\delta$, and the
coefficient $ \vert \Delta_g\vert^{k+2 }$ is modified (and
still positive). In some  examples the leading asymptotics is
provided by the minimal twists alone.

It is interesting to compare this with the twisted sectors. Since
the sector $(1,g)$ always mixes with $(g,1)$ under modular
transformation, and since the oscillator groundstate energy is
$-1$ in the untwisted sector,
 it is clear that for charges $Q$ corresponding to states in the
twisted sector the asymptotics will grow like
\eqn\twisted{ \hat I_{k+2 } (4\pi \sqrt{\half Q^2} ) }
This is true both for the absolute number of BPS states and for
the supertraces.
Recall that $k+2 = \half (n_v +2)$ for $\CN=2$ compactifications,
so we have   agreement with \iresulti.

There are some interesting general lessons we can draw from
our result \asympts.  Due to the factor $h(g)$   it is
possible that the leading $I$-Bessel functions cancel for certain
directions of $Q$. Moreover, a general feature of $\CN=2$
compactifications is that $g=1$ does not contribute to $\Omega_2$
in \asympts. Then, since $\vert \Delta_g \vert <1$ the
degeneracies are exponentially smaller in the untwisted sector
compared to those of the twisted sector.
 We will see an explicit example of this below.
In contrast, for $\CN=4$ compactifications, the $g=1$ term
{\it does} contribute to $\Omega_4$, which thus has the
same growth as in the twisted sector.

One general lesson seems
to be that the degeneracies, and even their leading asymptotics
can be sensitive functions of the ``direction'' of $Q$ in charge
space. In general it is quite possible that    the exact BPS degeneracies
and their asymptotics will be subtle arithmetic functions of the
charge vector $Q$.
\foot{Such a phenomenon was conjectured based on
other considerations in \MillerAG.}
In the physics literature it is often taken for granted that there
is a smooth function $S_n: H^{\rm even}(\CX,\IR) \to \IR$ so that
$S_n(sQ ) \sim \log \Omega_n(s Q)$ for $s\to \infty$, but the
true situation might actually be much more subtle.
The Rademacher expansion shows that Fourier coefficients of negative
weight modular forms have well-defined asymptotics governed by
Bessel functions. By contrast,
 the Fourier coefficients $a_n$ of   cusp forms of positive
weight $w$ have a lot of ``scatter'' and
can only be described by a probability distribution
for $a_n/n^{(w-1)/2}$. (See e.g.  \millergelbart\ for
an introduction to this subject.)
It would be very interesting to know where the functions
$\Omega_n(Q)$ fit into this dichotomy.

\newsec{Examples}

We now give some examples of the results one finds using these
general techniques. More details can be found in \DDMP.

\subsec{$ K3 \times T^2$}

This is dual to the heterotic string on $T^6$. We have
$\dim_{\CH_{BPS}(Q)} = p_{24}(N)$ where $N-1= \half Q^2$ and
$\eta^{-24} = q^{-1}\sum_{N=0}^\infty p_{24}(N) q^N$.
The Rademacher expansion (equation $(A.4)$ below)  becomes
\eqn\radptf{ \dim \CH_{BPS}(Q) = 16\cdot \Biggl[
\hat I_{13}(4\pi \sqrt{\half Q^2}) - 2^{-14} e^{i \pi \half
Q^2} \hat I_{13}(2\pi \sqrt{\half Q^2}) + \cdots \Biggr] }
For
$\Omega_4$ we simply replace $16$ by ${3\over 2}$. For $\Omega_6$ we find
\eqn\sxtwbx{ \Omega_6(Q) = {15\over 8} (2+ \half Q^2)
\hat I_{13}(4\pi \sqrt{\half Q^2}) + \cdots }
%
%%%%
and thus we conclude that the correct supertrace to use in \osvii\ is
$\Omega_4$, at least in this example.
%%%%
We thus see that - with a proper normalization of the measure
$d\phi$ - the integral expression \osvii\ agrees with the exact
degeneracies to {\it all orders in $1/Q^2$ in the leading
exponential.} We stress that this agreement arises just from using
the perturbative piece of $F(X^I, W^2)$. This is essentially the result of
\DabholkarYR. We also note that a naive inclusion of the worldsheet instanton
corrections does {\it not} lead to the subleading Bessel functions given
by the Rademacher expansion.

\subsec{ A reduced rank $\CN=4$ model}

Besides the simplest $K_3\times T^2$ compactification,
it is also possible to construct
a large number of $\CN=4$ type II models
by considering quotients of $K_3\times T^2$ by an Enriques
automorphism of $K_3$ combined with a translation on $T^2$.
We consider the simplest model
with $14$ $\CN=4$ vector multiplets,
corresponding to an Enriques involution with 8 odd two-cycles.
It is related by heterotic/type II duality \SchwarzBJ\ to
the $\bf Z_2$ orbifold of the $E_8 \times E_8$ string, where
the $\bf Z_2$ action interchanges the two $E_8$ factors and
simultaneously shifts half-way along a circle so that the twisted
states are massive \refs{\ChaudhuriFK,
\ChaudhuriBF}. The topological amplitude $F_1$ for this model
has been computed in \GregoriHI.

To apply the formalism of $\S{3}$, consider vectors
$(P_1,P_2,P_3,P_4)$ in $ E_8(-1) \oplus E_8(-1) \oplus II^{1,1}
\oplus II^{5,5}$ with orbifold action
\foot{The notation $E_8(a)$ used here and below means that the
$E_8$ lattice norm is scaled by an overall factor of $a$.}
\eqn\orbact{ g \vert
P_1,P_2,P_3,P_4\rangle = e^{2\pi i \delta \cdot P_3} \vert
P_2,P_1,P_3,P_4\rangle }
where   $2\delta \in II^{1,1}$ and $\delta^2=0$. The charge
lattice is $M_{el}= M_0 + M_1$ where $M_0$ are the charges of the
untwisted sector with
\eqn\eelat{ M_{0} = E_8(-\half ) \oplus II^{1,1} \oplus II^{5,5} }
while
\eqn\eelatp{ M_{1} = E_8(-\half ) \oplus (II^{1,1}+ \delta) \oplus
II^{5,5} }
are the charges in the twisted sector. For charges in the
untwisted sector we denote $Q= ({1\over \sqrt{2}} (2P+ \wp), P_3,
P_4)$ where $P\in E_8(+1)$, and $\wp$ runs over a set of lifts of
$E_8/2E_8$ to $E_8$. The absolute number of BPS states is given by
\eqn\thetnasin{ \dim\CH_{BPS}(Q) =  d^u_Q(N) }
for $N+ \Delta_{\wp} = \half
Q^2 $ where \eqn\dyou{
 8\Theta_{E_8(2), \wp}(\tau) {1\over \eta^{24} } +8  \delta_{\wp,0}
e^{2\pi i \delta \cdot P_3}
  {2^4\over \eta^{12} \vartheta_2^4} := q^{\Delta_\wp}
\sum_{N=0}^\infty d^u_Q(N) q^N }
with
\eqn\thetashf{ \Theta_{E_8(2), \wp}(\tau) := \sum_{Q\in E_8(+1)}
e^{2\pi i \tau( Q-\half \wp)^2 } }
The second supertrace vanishes, while for $\Omega_4$ we should multiply
by $3/32$. This expression only depends on $\wp$ up to the action
of the Weyl group of $E_8$. There are three orbits, of length
$1,120$ and $135$ corresponding to the trivial, adjoint, and
$3875$ representations. For each of these \thetashf\ may be
expressed in terms of theta functions.

For the twisted sector we define
\eqn\v{ \half \biggl(  {1\over \eta^{12} \vartheta_4^4} \pm
{1\over \eta^{12} \vartheta_3^4 } \biggr) = q^{\Delta_\pm} \sum_{N\geq 0}
d^t_{\pm}(N) q^N }
with $\Delta_+=-\half, \Delta_-=0$. The  absolute number of
twisted sector BPS states is given by
\eqn\twistabs{ \dim \CH_{BPS}(Q) = 16\cases{ d^t_+(N) & $e^{i \pi
Q^2} = -1$ \cr
 d^t_-(N) & $e^{i \pi Q^2} = +1$ \cr}
}
where $N+\Delta_\pm = \half Q^2$.

Applying the Rademacher expansion we find for $\Omega_{abs}(Q)=\dim \CH_{BPS}(Q)$:
\eqn\unstwasym{   \cases{\half
\hat I_9 (4\pi \sqrt{\half Q^2 } ) + 2^{-6}(15 + 16 e^{2\pi i P
\cdot \delta})  \hat I_9 (4\pi \sqrt{{1\over 4} Q^2 } )
+ \cdots &   $\vert \CO_{\wp}\vert = 1$ \cr
 \half    \hat I_9 (4\pi \sqrt{\half Q^2 } )
+ 2^{-6} \hat I_9 (4\pi \sqrt{{1\over 4} Q^2 } ) + \cdots
& $\vert \CO_{\wp}\vert = 120$ \cr \half   \hat I_9
(4\pi \sqrt{\half Q^2 } ) - 2^{-6}  \hat I_9 (4\pi
\sqrt{{1\over 4} Q^2 } ) + \cdots & $\vert \CO_{\wp}\vert = 135$
\cr
 \half  \hat  I_9\bigl(4\pi \sqrt{\half Q^2}\bigr)- 2^{-6} e^{i
\pi Q^2}   \hat I_9\bigl(4\pi \sqrt{{1\over 4}  Q^2}
\bigr) + \cdots & $Q\in M_1$ \cr} }
In the first three lines $Q\in M_0$ and $\vert \CO_{\wp}\vert$ is
the order of the $E_8$ Weyl group orbit of $\wp$. The leading term
is independent of the orbit, and in rather neat agreement with \iresult.

\subsec{The FHSV model}

As our third example let us consider the FHSV model. This has
$\CN=2$ supersymmetry
and is described in \FerraraYX. %
 We denote momentum  vectors by $(P_1,P_2,P_3,P_4)$ in
$II^{9,1} \oplus II^{9,1} \oplus II^{1,1} \oplus II^{3,3}$ The
$\IZ_2$ acts as
\eqn\ztwoact{
 \vert P_1, P_2, P_3, P_4 \rangle
\rightarrow e^{ 2\pi i \delta \cdot P_3} \vert P_2, P_1, P_3, -P_4
\rangle }
 with $\delta$ the
order two shift vector defined in \FerraraYX ($\delta^2 = \half $).
The $u(1)^{12}$ electric charge lattice is
 $M_{el} = M_{0} + M_{1}  $
where
\eqn\eelat{ M_{0} = E_8(-\half ) \oplus II^{1,1}(\half)  \oplus
II^{1,1} }
\eqn\eelatp{ M_{1} = E_8(-\half ) \oplus II^{1,1}(\half) \oplus
(II^{1,1}+\delta) }
States from the untwisted sector have charge vectors in $M_0$,
while states from the twisted sector have charge vectors in $M_1$.

In order to give the degeneracies of DH states we define
\eqn\radmachi{ \eqalign{
&  {2^6\over \eta^6 \vartheta_2^6} =
q^{-1} \sum_{N=0}^{\infty} d^u(N) q^N \cr
& \half \biggl(  {1\over \eta^{6}
\vartheta_4^6} +   {1\over \eta^{6} \vartheta_3^6 } \biggr) =
q^{-{1\over 4}} \sum_{N=0}^\infty d^t_+(N) q^N\cr
& \half
\biggl(  {1\over \eta^{6} \vartheta_4^6} -   {1\over \eta^{6}
\vartheta_3^6 } \biggr) = q^{+{1\over 4}} \sum_{N=0}^\infty
d^t_-(N) q^N\cr} }
Then, for  the helicity supertrace in the untwisted sector   we have the result:
\eqn\btwores{ \Omega_2(Q) = \cases{   e^{2\pi i Q\delta}
d^u(N)& $Q\in M_0'$ \cr 0 & $Q\in M_0-M_0' $\cr} }
where $N-1 = \half Q^2$ and $M_0'$ is the sublattice of   vectors
of the form $2P_1 \oplus 2 P_2 \oplus P_3$ of $M_0$.

For the twisted sector, note that $Q\in M_1$ and hence $Q^2 \in
\IZ+{1\over 2}$ The exact second supertrace is
\eqn\polus{ \Omega_2(Q) = \cases{ -16 d_+^t(N) \qquad & for $e^{i
\pi Q^2}   = - i$ \cr -16 d_-^t(N)   &  for $ e^{i \pi Q^2}  = +i$
\cr} }
The oscillator level $N$ is related to the momentum by the
condition $N+\Delta_\pm = \half Q^2$
and the $\pm$ sign is correlated  with the sign of \polus. Note
that the metric $II^{9,1}(\half) \oplus (II^{1,1}+\delta)$ is used
here.

Using the Rademacher expansion we have the asymptotics
\eqn\asympomegatwo{ \Omega_2(Q) = \cases{  2^{-8} e^{2\pi i Q\cdot
\delta} (1- e^{i \pi Q^2/2} )  \hat I_7(2\pi \sqrt{\half Q^2}) +
\CO(e^{\pi \sqrt{Q^2/2}}) & $Q\in M_0' $ \cr 0 & $Q\in M_0 -
M_0'$ \cr - 2^{-3} \hat I_7(4\pi \sqrt{\half Q^2} ) +
2^{-11} i e^{i \pi Q^2}  \hat I_7(2\pi \sqrt{\half Q^2} ) +
\CO(e^{\pi \sqrt{Q^2/2}})
 & $Q\in M_1$ \cr} }

Let us now compare these results with \iresult\iresulti\ and hence with
\osvii\modosv. The degeneracies in the twisted sector are consistent with
\iresult\ but this does not appear to be the case for the
untwisted sector, because the exponential growth is $\exp[2\pi \sqrt{\half Q^2}]$.
\foot{This discrepancy is avoided in a class of $\CN=2$
heterotic orbifolds where twisted states carry the same charges
as untwisted states, hence dominate the helicity supertrace \DDMP.
In the FHSV model, twisted and untwisted states can be distinguished by
the moding of the winding number.}
It is interesting also to consider the absolute number of
BPS states in the untwisted sector.  These are given by
$\dim \CH_{BPS}(Q) = \alpha(N )$ where $N-1 =\half Q^2$ and
\eqn\aboslt{
{8\over \eta^{24}} \CF_{g,Q}(q) = q^{-1}\sum_{N\geq 0} \alpha(N) q^N
 }
For generic moduli, the asymptotics of the
absolute number of BPS states is controlled by $\hat I_{13}(4\pi
\sqrt{\half Q^2})$. However $\CF_{g,Q}$ is a function of moduli
and on  some subvarieties of moduli space $\CF_{g,Q}$ can be
enhanced to an $E_8$ theta function. In this case the absolute
number of BPS states is enhanced to $\hat I_{9}(4\pi \sqrt{\half
Q^2})$. Thus, the leading exponential behavior is the desired
$\exp[4\pi \sqrt{\half Q^2}]$ but the logarithmic corrections are
in fact moduli-dependent. This is to be contrasted with the
supertrace $\Omega_2$, which is moduli independent, but
for $Q\in M_0'$ goes like $\hat I_7(2\pi \sqrt{\half Q^2})$, and
is exponentially smaller than the absolute number of BPS states.

Thus the exact degeneracies do not agree with \osvii\modosv\ with
any natural interpretation of $\Omega$. However, as explained in
 $\S{2.4.4}$  the  integrals  \osvii\modosv\ are highly
singular. Thus the formalism of \OoguriZV\ breaks down and
this discrepancy cannot be said to  constitute a counterexample to the conjecture of
\OoguriZV.

\subsec{Purely electric states}

It is also instructive to consider purely electric states, i.e.
those with $p^a=0$ but
$q_a\not=0$.
An interesting example where such states can be investigated in
detail are the perturbative
type II DH states in $K3 \times T^2$ compactification. These
states are   obtained from  fundamental type II strings with
momentum and winding along the $T^2$ factor. These are purely electric
states in the natural polarization for the type II string. They are
related by $U$-duality to BPS states of $D2$ branes wrapping a $T^2$ and
a holomorphic curve  in the
$K3$ surface. In this case $p^a=0$, so that the perturbative
part of the free energy \cfhptv\ vanishes, while the exact
free energy is given by
\eqn\cfsu{
\CF(\phi,p) = - \log \vert \Delta(\tau)\vert^2
}
for $\tau=\phi^1/\phi^0$. As a consequence, 
the integral \osvii\ is highly singular.
Nevertheless we have (see  \KiritsisHJ, eqs. $(G.24)$ and $(G.25)$):
\eqn\typestii{
\eqalign{
\Omega_4(Q) & = 36\ \delta_{Q^2,0} \cr
\Omega_6(Q) & = 90\ \delta_{Q^2,0}  \cr}
}
for charges $Q$ such as we have described. Meanwhile
$\Omega_{abs}(Q)$ grows exponentially, like $\exp[2\pi \sqrt{Q^2}]$.
Note that in contrast to the heterotic case,  for 
$Q^2\not=0$ these states are ${1\over 4}$-BPS, despite the 
fact that their discriminant vanishes.   Further discussion of these
states, and related states in type $(0,4)$/$(2,2)$ duality pairs will be
given in \DDMP.

\subsec{Large black holes and the $(0,4)$ CFT dual}

Regrettably, there
 are no examples where the degeneracies of large black holes
are known exactly. In principle the index $\Omega_2$ should be
computable from a $(0,4)$ sigma model described in
\MaldacenaDE\MinasianQN, presumably from the elliptic genus
of this model. While the sigma model is rather complicated,
and has not been well investigated we should note that
from the Rademacher expansion it is clear that the
leading exponential asymptotics of negative weight modular forms
depends on very little
data. Essentially all that enters is the order of the pole and 
the negative modular weight. There are $c_L =   C(p) + c_2\cdot p= \hat C(p)$ real left-moving
bosons. Since the sigma model is unitary, the
relevant modular form has the expansion
$q^{-c_L/24} + \cdots $. This gives the order of the pole, and thus 
we need only know the modular weight. This in turn depends on 
the  number of left-moving noncompact bosons. Each noncompact boson 
contributes $w=-\half $ to the modular weight. Now, the sigma model 
of \MaldacenaDE\ splits into a product of a relatively simple 
``universal factor'' and a rather complicated ``entropic factor,'' 
as described in \MinasianQN. Little is known about the entropic 
factor other than that it is a $(0,4)$ conformal theory with 
$c_R = 6 k$, where $k= {1\over 6} C(p) + {1\over 12} c_2\cdot p-1$, where $p\in H^2(\CZ,\IZ)$. 
The local geometry of the target space was worked out in \MinasianQN. 
Based on this picture we will  assume the target space is 
compact and does not contribute to the modular weight. 
(Quite possibly the model is a ``singular conformal field theory'' 
in the sense of \SeibergXZ\ because the surface in the linear 
system $\vert p\vert$ can degenerate along the discriminant locus. 
It is reasonable to model this degeneration using a Liouville 
theory, as in \SeibergXZ. If this is the case we expect   the entropic 
factor to contribute order one modular weight.) 
The universal factor is much more explicit. The target is $\IR^3 \times S^1$, 
it has $(0,4)$ supersymmetry with $k=1$ and there are $h-1$ compact leftmoving 
bosons which are $N=4$ singlets. They have momentum in the 
anti-selfdual part of $H^{1,1}(\CX,\IZ)$ (anti-selfduality is 
defined by the surface in $\vert p \vert$). Since we fix these 
momenta we obtain $w=-\half (h-1)$. Finally there are 3 
noncompact left-moving bosons 
describing the center of mass of the black hole in $\IR^3$. Thus, the 
net left-moving modular weight is $-(h+2)/2$. Now, applying the 
Rademacher   expansion in the region $\vert \hat q_0 \vert \gg \hat C(p)$
we find the elliptic genus is proportional to 
\eqn\radmsw{
\hat I_{\nu}\biggl(2\pi \sqrt{{\vert \hat q_0 \vert \hat C(p)\over 6}}\biggr)
}
with $\nu =  {h+4\over 2}$. This is remarkably close to  \ibessl !  
Clearly, further work is needed here since it is likely there are 
a number of important subtleties in the entropic factor. Nevertheless,  
 our argument suggests that a deeper investigation of the elliptic 
genus in this model will lead to an interesting test of  
 \osvii\ (or rather \modosv, since it must be
done at strong topological string coupling) for the case of large black holes.

\newsec{Conclusions}

We have seen that the heterotic DH states and the corresponding
small black holes provide a rich set of examples for testing the
precise meaning and the range of validity of  \osvii. We have
computed exactly  the absolute number of DH states in a large
class of  orbifold compactifications with  $\CN =4$ and $\CN =2$
supersymmetry. We have also evaluated various supertraces which
effectively count the number of `unpaired' BPS short multiplets
that do not have the spin content to combine into long multiplets.
These supertraces provide valuable information about how the BPS
spectrum is organized and are important for finding the correct
interpretation of our results. Using these data, a far more
detailed comparison of microscopic and macroscopic degeneracies
can be carried out than is possible for large black holes. We
summarize below our results along with a number of puzzles and open
problems and conclude with possible interpretations.

\subsec{Results}

On the macroscopic side, the asymptotic black hole degeneracies
are proportional  to a Bessel function \ibessl\iresult. For heterotic DH
states with a charge vector $Q$, the Bessel function is of the
form $\hat I_\nu(4\pi\sqrt{Q^2/2})$ where the index $\nu$ is given in
terms of the number of massless vector fields by \iresulti.
If instead one considers a limit of charges with weak topological
string coupling and $\chi(\CX)\not=0$ then the asymptotics are far more
more complicated than those of a Bessel function, and are given by
\contribent\ , in leading order.

On the microscopic side, the absolute number of the untwisted DH
states is given by the general formulae \bsnttn, \andthcz. The
asymptotic microscopic degeneracies of the untwisted states are
given by \asympts\ and of the twisted states by \twisted. These
are both expressed in terms of an $I$-Bessel function. Asymptotically, the
relevant supertraces are also Bessel functions. All these Bessel
functions in general have different arguments and indices.

Comparison of these asymptotic degeneracies reveals the following
broad patterns which we have checked in a few explicit examples
here and many other examples that will be reported in \DDMP.

$\bullet$ In all reduced rank CHL-type orbifolds with $\CN=4$
supersymmetry,  there is  remarkable agreement between the
microscopic and macroscopic degeneracies for all possible charge
vectors in both twisted and untwisted sectors. See for example
\unstwasym.   The agreement holds to all orders in an asymptotic
expansion in $1/Q^2$, but fails nonperturbatively.
\foot{Nonperturbative discrepancies in the formula \osvii\
have previously been addressed in \VafaQA\AganagicJS. The
systems discussed in these papers are very different from the
compact Calabi-Yau case discussed in this paper.}
It is
noteworthy that this agreement uses only the perturbative part of
the topological string partition function and worldsheet
instantons play no role.

The relevant helicity supertrace in this case is $\Omega_4$ which
turns out to be proportional to the absolute number because the
left-moving oscillators of the heterotic string do not carry any
spacetime fermion numbers, so there are no intermediate BPS representations.

$\bullet$ In orbifolds with $\CN=2$ supersymmetry,  the leading
order microscopic  entropy is determined entirely
by the argument of the Bessel function and in all models it goes
as $4\pi \sqrt{Q^2/2}$.  This is expected from a general argument
in \SenIS\ that if the entropies match in the toroidally
compactified heterotic string, as they do \DabholkarYR, then they
must also match  in all $\CN=2$ orbifolds. The
subleading terms however depend also on the index of the Bessel
function and these match only for twisted states but not for the
untwisted states. The relevant nonvanishing helicity supertrace in
this case is $\Omega_2$. For the twisted states, $\Omega_2$ is
proportional to the absolute number. For the untwisted states,
$\Omega_2$ is exponentially smaller than the absolute number
because the argument of the corresponding Bessel function turns
out to be $2\pi \sqrt{Q^2/2}$ and moreover the index is also
different.

Unfortunately, as we have explained in $\S{2.4.4}$
in this case we cannot reliably compute the macroscopic
degeneracy because the prescription in \OoguriZV\ forces us to work
on the boundary of Teichm\"uller space, and $F_{\rm top}$ is singular
on this locus. Nevertheless, remarkably, if we ignore this
subtlety and consider the result \iresult\ we find
precise agreement for the twisted sector DH states. We
find disagreement both with $\Omega_{abs}$ and with $\Omega_2$
for the untwisted sector DH states.

$\bullet$ We have focused in this paper on the heterotic DH
states, but it is instructive to consider also the Type-II DH
states, as discussed in sec. 4.4.
%
% For example, in  $K3 \times T^2$ compactification of the
%Type-II string, these states are obtained from fundamental Type-II
%strings with momentum and winding along the $T^2$ factor. These
%are purely electric states in the natural polarization for the
%type II string. They are related by $U$-duality to BPS states of
%$D2$ branes $T^2$ and a holomorphic curve  in the $K3$ surface.
%
In this case, since $p^I =0$, the graviphoton charge vanishes
and the integral \osvii\ becomes quite singular, even in cases
where the exact $F_{\rm top}$ is known.  Moreover,
 even after the inclusion of the F-type terms, the
geometry continues to have a null singularity and does not develop
a regular horizon. It is not clear in this case how to apply the
formalism implicit in \osvii\ and it is likely that the D-type
terms are important for desingularizing these solutions. These
states will be discussed in more detail in \DDMP.

\subsec{Puzzles and open problems}

Our results raise a number of questions and puzzles. Their
resolution is essential for a correct interpretation of \osvii.

$\bullet$ An important assumption underlying \osvii\ both for the
large and small black holes is that the D-type terms  in the low
energy effective action do not contribute to the black hole
entropy. A priori, it is far from clear if that is the case.

The strikingly successful agreement for the large class of
heterotic DH states in $\CN=4$ orbifolds strongly suggests that at
least for this class of small black holes, the D-terms in fact do
not modify the entropy. It is highly unlikely that various precise
numerical factors could have come out right only accidentally. It
is quite conceivable for instance that once the F-type quantum
corrections generate a solution with a regular horizon, then  on
that background solution,  the corrections from the D-type terms
do not change the Wald entropy possibly because of the index
structure of the background Riemann tensor and gauge fields.
There are analogous situations where a similar phenomenon occurs,
for example, in $AdS_5 \times S^5$ or in chiral null models, where
the higher curvature terms do not alter the solution because of
the specific details of the index structure. It would be very
interesting to see explicitly if this is indeed the case for our
small black holes.

The Type-II DH states noted in the previous subsection
also suggest that in general, the D-type terms {\it
will} be important. In this case, the F-type terms are inadequate
to desingularize the solution. Following the heuristic picture of
the stretched horizon suggested in \SenIN,  one is then forced to
include the D-type terms to obtain a solution with a regular
horizon to be able to make a meaningful comparison with the
microstates.
This suggests that even for large black holes, whether or not the
effect of D-type terms needs to be included may depend on the
details of the model and on the class of states.

$\bullet$
We have seen that even in the successful cases, \osvii\ (or rather,
the more accurate \modosv)
is only true in perturbation theory. If one wishes to go beyond
the asymptotic expansion and understand \osvii\ as a statement
about exact BPS degeneracies, then one must specify a
nonperturbative definition of $\psi_p$ and must then specify
carefully the region of integration. Regarding the first problem,
the $K3 \times T^2$ example is of fundamental importance because
the   $K3 \times T^2$ wavefunction is known exactly. In this case
we can say definitively that $\psi_p$ is not a normalizable
wavefunction and therefore not in the Hilbert space \DDMP. It is
important and interesting to investigate this issue for other
Calabi-Yau manifolds, but without a nonperturbative definition it
is impossible to make definitive statements. Nevertheless, in the
examples of $\CX$ with heterotic duals, the functions $F_g$   are
automorphic functions of the $t^a$. See, for examples,
\refs{\DixonPC,\deWitZG,\AntoniadisCT, \HarveyFQ, \MooreAR,
\MarinoPG, \KlemmKM}. This is already sufficient knowledge to
address to some extent the question of what contour of integration
should be chosen for the $\phi^I$.
%%%%
We have seen that if   we keep just the perturbative part of $F$
then it is natural to  integrate $\phi^I$ along the imaginary
axis.
%%%%
However, this is problematic if we wish to retain the worldsheet
instanton corrections. When $t^a:=X^a/X^0$ has a  positive
imaginary part the instanton series in \largrad\  at fixed $g$,
but summed over $\beta$ converges.  Automorphic  forms are highly
singular when evaluated for $t^a$ purely real. This can already be
seen in the $ K3 \times T^2$ example, where one is evaluating
$\Delta(\tau)$ for real $\tau$. If one tries instead to expand the
integrand of \osvii\ using the expansion in Gromov-Witten
invariants
 one finds an infinite series of order one terms leading to a
nonsensical result. (In particular, the expansion in worldsheet
instantons does not lead to the subleading exponentials in the
Rademacher expansion.)
%%%%

How then are we to understand \osvii? One possibility is that the
full nonperturbative topological string partition function defines
an $n$-form $\omega_p = d\phi e^{\CF}$ with singularities on
$H^{even}(\CX,\IC)$ and that certain periods of this form give
$\Omega(p,q)$. Then our procedure above could be a saddle point
approximation to such a contour integral, and the Bessel functions
\ibessl\iresult\ represent the full asymptotic expansion
multiplying the leading exponential. At least this interpretation
is consistent with the data provided by perturbative heterotic
states.

$\bullet$
An interesting question raised by the subleading Bessel functions
in the Rademacher expansion is that of their physical meaning. The
subleading corrections to $p_{24}(N)$ in the case of $K3 \times
T^2$ are down by $exp[-4\pi {c-1\over c} \sqrt{N}]$,
$c=2,3,\dots$,  and since $\sqrt{N} \sim 1/g_s^2$ at the horizon
this is suggestive of some novel nonperturbative effects.

\subsec{Interpretations}

One interpretation that has been suggested in \OoguriZV\ 
is that the quantity $\Omega$ appearing in
\osvii\ is not the absolute number of micro-states but rather an index. It is
natural to identify this proposed index with $\Omega_4$ (or $\Omega_6$)
 in $\CN=4$ theories and with $\Omega_2$ in $\CN = 2$ theories.

In all successful examples where the agreement works,  this index
always equals the absolute number and also the macroscopic black
hole degeneracy. This seems to support the above interpretation.
However,
%the untwisted states in the $\CN=2$ orbifolds provide an
%explicit counterexample. For these states, the index is
%exponentially smaller than  the absolute number because a large
%number of vector-type short multiplets and hyper-type short
%multiplets cancel in pairs. The absolute number agrees with the
%black hole degeneracy to leading order because the argument of the
%Bessel functions agrees,  but not with  the sub-leading orders because
%the index of the Bessel function does not agree.
%
the interpretation in terms of an index   seems problematic from the point of view
of thermodynamics. The Bekenstein-Hawking-Wald entropy appears in
the first law of thermodynamics which can be derived in the
Lorentzian theory where there are no ambiguities about fermionic
boundary conditions. As with any other thermodynamic system,  one
should identify this entropy with the logarithm of the absolute
number of microstates by the Boltzmann relation and not with an
index. Generically, the index will be much smaller than the
absolute number because many states can cancel in pairs when
counted in an index and thus cannot equal the thermodynamic
entropy. This problem is even more acute for large black holes. In
this case, the classical area is finite and any possible quantum
corrections due to the F-type and D-type terms are subleading. On
general grounds, it does not seem reasonable to identify this
thermodynamic entropy with an index.

Our results suggest a possible alternative
interpretation that the macroscopic entropy should be compared
with the absolute microscopic degeneracies, but that these
degeneracies must be computed in an appropriate ``nonperturbative'' regime of
moduli space. Indeed this is what one would expect from the Boltzmann
relation in conventional statistical mechanics.
\foot{In fact, not only $\Omega_{abs}$ but also $\Omega_2$ is only a
locally constant function on moduli space. The function $\Omega_2$
can  change across walls of marginal stability in vectormultiplet
moduli space (although it is constant in hypermultiplet moduli space).
Thus, even a version of \osvii\ in which $\Omega$ is given by
an index must  also take into
account the region of moduli space in which $\Omega$ is being computed.}

Note that even if the string coupling remains small at the
horizon it does not mean that we are in a perturbative regime
because the graviphoton charge of the state of interest has to be
large enough so that a black hole is formed. Formation of a black
hole is clearly a nonperturbative change in  the perturbative flat
spacetime geometry. This is analogous to a situation in QED where
even if the fine structure constant $\alpha$ is small, the
interactions of a particle with charge $Z$ cannot be computed in
perturbation theory  for sufficiently large $Z$ once $\alpha Z$ is
of order one.

Therefore, for a correct comparison, we need to evaluate the
microscopic degeneracies in the regions of the moduli space
determined by the attractor geometry where a black hole has
formed. We are instead computing the microscopic degeneracies in
the perturbative regime using free string theory in flat
spacetime. The two computations do not always have to agree even
for BPS states in short multiplets because with the right spin
content, many short multiplets can in principle combine into a
long multiplet. The long multiplets are then not protected from
renormalization. This suggests that the spectrum of BPS short
multiplets would be robust against renormalization only when their
absolute number equals an index and that index is itself constant.
In this case, the short
multiplets  cannot turn into a long multiplet because they simply
do not have the required spin content.

This interpretation is indeed consistent with our results for all
heterotic DH states. Whenever the perturbative microscopic
degeneracies match with macroscopic degeneracies as in the $\CN
=4$ models or for the twisted states in the $\CN =2$ models, they
also equal an index. It seems reasonable to expect that in this
case the microscopic degeneracies in the nonperturbative black
hole regime can be reliably deduced from the microscopic
degeneracies in the perturbative regime.

%Finally, it should be kept in mind that the conjectural relation
%\osvii\  between
%microscopic degeneracies and the topological string amplitude is logically
%disconnected from the relation between the topological string amplitude
%and the macroscopic Bekenstein-Hawking-Wald entropy. It may well be that
%some version of the former hold for general BPS black holes, while the
%latter may be valid for a restricted class of black holes only. In this
%case, it is clear that $\Omega(p,q)$ cannot refer to absolute
%degeneracies, but only to helicity supertraces.
%

\bigskip
\noindent{\bf Acknowledgements:} It is a pleasure to thank J.
Kappeli, E. Kiritsis, C. Kounnas, S. Metzger, S. Miller, H. Ooguri,  N. Saulina,  A. Sen, A.
Strominger, C. Vafa, P. Vanhove, and E. Verlinde
for valuable discussions. We thank A. Sen for sending us a draft of
\SenPU\ prior to publication.  We also thank A. Strominger
for useful remarks on a draft. The work of FD and GM is
supported in part by DOE grant DE-FG02-96ER40949.
The work of BP is supported in part by the
European Research Network
MRTN-CT-2004-512194.

\bigskip\bigskip
\noindent
{\it Note added}: Both versions 1 and 2 of this paper asserted that
the degeneracies of untwisted DH states in $\CN=2$ orbifold compactifications
constituted a counterexample to the conjecture of \OoguriZV.
We subsequently realized that in these examples our computation 
of the integral \osvii\ in $\S{2.3}$ is not rigorous because 
certain K\"ahler classes are zero at the attractor point.
\foot{We disagree with the statement in  footnote 2 of \DijkgraafBP.
In fact, for the FHSV example the computation can be done at weak
coupling. In particular, the nonperturbative effects discussed in
\DijkgraafBP\ , of order $\CO(e^{-t^2/\lambda})$, are exponentially
small in the limit we consider.}
For further explanation and discussion see   $\S{2.4.4}$ and
$\S{5.1}$. In the present revised version, our claims 
are requalified as follows: we find rigorous agreement for $\CN=4$ 
compactifications, remarkable 
{\it unjustified} agreement for twisted sector $\CN=2$ DH states, and
{\it apparent} discrepancy for untwisted $\CN=2$ DH states.
In fact, the formula \osvii\ appears to be rather singular in this case.
We have also taken the opportunity to add some
new results in $\S{2.4.3}$ and $\S{4.5}$.

\bigskip

\appendix{A}{The Rademacher expansion}

Here we state briefly the Rademacher expansion. For more details
and information see \DijkgraafFQ.

Suppose  we have a ``vector-valued nearly holomorphic modular
form,''  i.e.,  a collection of functions $f_\mu(\tau)$ which form
a finite-dimensional unitary representation of the modular group
of weight $w<0$. Under the standard generators we have \eqn\rep{
\eqalign{ f_\mu(\tau+1) & = e^{2\pi i \Delta_\mu} f_\mu(\tau)\cr
f_\mu(-1/\tau) & = (-i \tau)^w S_{\mu\nu} f_\nu(\tau) \cr} }

We assume the $f_\mu(\tau)$ have
 no singularities for $\tau$ in the upper half plane,
except at the cusps $\IQ \cup i \infty$. We may assume they have
an absolutely convergent Fourier expansion \eqn\collec{
f_\mu(\tau) = q^{\Delta_\mu} \sum_{m \geq 0} F_\mu(m) q^m \qquad
\mu = 1, \dots, r } with $F_\mu(0)\not=0$ and that the
$\Delta_\mu$ are real.  We wish to give a formula for the Fourier
coefficients $F_\mu(m)$.

Define: \eqn\integrl{ \hat I_\nu(z) =  -i (2\pi)^{\nu}  \int_{\epsilon-i\infty}^{\epsilon+i\infty} t^{-\nu-1}
 e^{(t + z^2/(4t))} dt = 2\pi ({z \over  4\pi} )^{-\nu} I_\nu(z)
} for $Re(\nu)>0, \epsilon>0$,
where $I_\nu(z)$ is the standard modified Bessel
function of the first kind.

Then we have:
 \eqn\radi{ \eqalign{ F_\nu(n) =  \sum_{c=1}^\infty\sum_{\mu=1}^{r} &
  c^{w-2} K\ell(n,\nu,m,\mu;c)
\sum_{m+ \Delta_\mu < 0} F_\mu(m)\cr \vert m+\Delta_\mu
\vert^{1-w} & \hat I_{1-w}
 \biggl[ {4\pi\over c} \sqrt{\vert m+\Delta_\mu\vert(n + \Delta_\nu)}
\biggr] . \cr} }
The coefficients $ K\ell(n,\nu,m,\mu;c)$ are
generalized Kloosterman sums. For $c=1$  we have: \eqn\radiii{
K\ell(n,\nu,m,\mu;c=1)   = S^{-1}_{\nu\mu} } The series \radi\ is
convergent. Moreover the asymptotics of $I_\nu$ for large $Re(z)$
is given by \eqn\inuas{
 I_\nu(z) \sim \frac{e^z}{\sqrt{2\pi z}}\left[ 1- \frac{(\mu
    -1)}{8z} + \frac{(\mu
    -1)(\mu -3^2)}{2!(8z)^2} - \frac{(\mu
    -1)(\mu -3^2)(\mu -5^2)}{3!(8z)^3}+ \ldots \right],
} where $\mu = 4 \nu^2$.

\listrefs

\bye